# 6. ARTIGO

**Equalização das escalas NESSCA e SARA utilizando a Teoria da Resposta ao Item na avaliação do comprometimento pela doença de Machado-Joseph**

Nicole Machado Utpott[1], Vanessa Bielefeldt Leotti[1,2], Laura Bannach Jardim[3,4]

[1] Programa de Pós-Graduação em Epidemiologia, Faculdade de Medicina, Universidade Federal do Rio Grande do Sul, Porto Alegre, Brasil

[2] Departamento de Estatística, Instituto de Matemática e Estatística, Universidade Federal do Rio Grande do Sul, Porto Alegre, Brasil

[3] Programa de Pós-Graduação em Ciências Médicas, Faculdade de Medicina, Universidade Federal do Rio Grande do Sul, Brasil

[4] Serviço de Genética Médica, Hospital de Clínicas de Porto Alegre, Brasil

**RESUMO**

**Contexto:** A equalização de escalas é uma técnica estatística utilizada para estabelecer relações de equivalência entre diferentes escalas. Sua utilização é bastante popular na avaliação educacional, porém, incomum na área da saúde, onde escalas de medidas são ferramentas integrantes da prática clínica. Com a utilização de diferentes escalas, existe uma dificuldade em comparar resultados científicos, como é o caso das escalas NESSCA e SARA, instrumentos de avaliação do comprometimento pela doença de Machado-Joseph (SCA3/MJD). **Objetivo:** Explorar o método de equalização de escalas e demonstrar sua aplicação através das escalas NESSCA e SARA, utilizando a abordagem da Teoria da Resposta ao Item (TRI) na avaliação do comprometimento pela SCA3/MJD. **Métodos:** Os dados são de 227 pacientes do Hospital de Clínicas de Porto Alegre portadores da SCA3/MJD que possuem medidas completas para NESSCA e/ou SARA. O delineamento de equalização utilizado é o de grupos não equivalentes com itens comuns, com calibração separada. O modelo TRI utilizado na estimação dos parâmetros foi o de crédito parcial generalizado, para NESSCA e SARA. Foi feita a transformação linear através dos métodos *Mean/Mean*, *Mean/Sigma*, *Haebara* e *Stoking-Lord* e foi aplicado o método da equalização do verdadeiro escore para obter uma relação estimada entre os escores das escalas. **Resultados:** O escore NESSCA estimado pela equalização via escore SARA comparado com o escore NESSCA observado apresentou diferença mediana de 0,82 pontos pelo método *Mean/Sigma*. Este foi o melhor método de transformação linear dentre os testados. **Conclusões:** Com este estudo foi possível explorar a aplicabilidade da técnica de equalização via TRI no contexto da saúde e ilustrar sua utilização criando uma relação de equivalência entre os escores das escalas NESSCA e SARA.

**Palavras-Chave:** Equalização de escalas; Teoria da Resposta ao Item; doença de Machado-Joseph; NESSCA; SARA.

## ABSTRACT

**Background:** Scale equating is a statistical technique used to establish equivalence relations between different scales. Its use is quite popular in educational evaluation, however, unusual in the health area, where scales of measures are tools that integrate clinical practice. With the use of different scales, there is a difficulty in comparing scientific results, such as NESSCA and SARA scales, tools for assessing the commitment to Machado-Joseph disease (SCA3/MJD). **Objective:** Explore the method of scale equating and demonstrate its application through NESSCA and SARA scales, using the Item Response Theory (IRT) approach in assessing SCA3/MJD commitment. **Methods:** Data came from 227 patients from the Hospital de Clínicas de Porto Alegre with SCA3/MJD who have complete measures for NESSCA and/or SARA scales. The equating design used is that of non-equivalent groups with common items, with separate calibration. The IRT model used in the estimation of the parameters was the generalized partial credit, for NESSCA and SARA. The linear transformation was performed using the Mean/Mean, Mean/Sigma, Haebara and Stoking-Lord methods and the equation of the true score was applied to obtain an estimated relationship between the scores of the scales. **Results:** Difference between NESSCA score estimated by SARA and observed NESSCA score has shown median of 0.82 points, by Mean/Sigma method. This was the best method of linear transformation among the tested. **Conclusions:** This study extended the use of scale equating under IRT approach to health outcomes and established an equivalence relationship between NESSCA and SARA scores, making the comparison between patients and scientific results feasible.

**Key Words:** Scale equating; Item Response Theory; Machado-Joseph disease; NESSCA; SARA.

## INTRODUÇÃO

A equalização de escalas é uma técnica estatística utilizada para estabelecer relações de equivalência entre diferentes escalas (1). Embora a área da saúde seja uma das áreas de conhecimento que mais desenvolvem e consomem questionários e testes, a técnica de equalização ainda é pouco explorada neste contexto (2). Sua base teórica é fundamentada na psicometria e sua utilização é majoritariamente voltada para a avaliação educacional (2,3). Na área da saúde, as escalas de medidas são ferramentas integrantes da prática clínica e da avaliação do estado de saúde do paciente, pois possuem grande influência sobre as decisões que se referem ao tratamento e que envolvem a gestão de políticas públicas direcionadas para o cuidado com a saúde da população. O avanço em pesquisas e as necessidades cada vez mais específicas dos pacientes contribuem para o surgimento de muitas ferramentas de medida (4). Nesse sentido, a equalização torna possível a criação de uma relação entre diferentes escalas, possibilitando a comparação entre resultados de publicações científicas.

Tradicionalmente, as escalas de medidas são analisadas de acordo com os princípios da Teoria Clássica dos Testes (TCT), que se baseia na média ou na soma dos escores obtido nos itens. A Teoria da Resposta ao Item (TRI), tecnicamente mais robusta, foi desenvolvida para suprir as principais limitações da TCT, como a ausência de discriminação entre itens, ou seja, respondentes com o mesmo escore total são considerados iguais mesmo que o conjunto de respostas tenha sido totalmente diferente (1,5). A TRI tem por objetivo descrever a associação entre a probabilidade de uma resposta a um item em particular e o nível de um respondente quanto a uma característica de interesse que não pode ser observada diretamente, conhecida por traço latente (5). O estado de saúde de um paciente é uma variável que, na estatística, pode ser classificada como um traço latente. Apesar de não poder ser medido diretamente, o traço latente pode ser inferido com base na observação de variáveis secundárias que estejam relacionadas a essa característica de interesse e que possam ser mensuradas através de instrumentos de medidas (6). Os modelos TRI variam conforme a natureza do item (dicotômicos ou politômicos), o número de populações envolvidas e a quantidade de traços latentes avaliados (1). A escolha do modelo deve ser adequada às escalas.

Um exemplo de patologia que utiliza escalas para mensurar o estado de saúde do paciente é a doença de Machado-Joseph. Também conhecida como ataxia espinocerebelar tipo 3 (SCA3/MJD), é um distúrbio neurodegenerativo autossômico dominante caracterizado por uma ataxia cerebelar de início geralmente na idade adulta (7). Apesar de não existir cura para a doença, existem algumas opções para administrar os sintomas (8). Para isso, a avaliação do nível de comprometimento neurológico dos pacientes é essencial. Em 2006, foi publicada a *Scale for Assessment and Rating of Ataxia* (SARA) (9) que, por sua simplicidade, tem sido bastante utilizada nas publicações científicas. A escala

SARA é composta por 8 itens, cada um tem de cinco a nove categorias de resposta, cujo escore total varia de 0 a 40 pontos. Em 2008 foi publicada a *Neurological Examination escore for Spinocerebellar Ataxia* (NESSCA) (10), que diferencia-se da SARA por avaliar também sintomas não-atáxicos. A NESSCA é dividida em 18 itens, cada um possui de duas a cinco categoriais de resposta, somando, no total 40 pontos. Tradicionalmente, as escalas SARA e NESSCA estimam o traço latente considerando a soma do escore obtido em cada item, que é o pressuposto da TCT.

No contexto da SCA3/MJD, a TRI modela a relação existente entre a probabilidade de um paciente apresentar um sintoma e o nível do comprometimento da SCA3/MJD. Em 2013 a NESSCA foi avaliada através dos modelos da TRI com dados de 106 pacientes. Os autores identificaram os sintomas que ajudam a explicar melhor o comprometimento com a doença e propuseram alterações na escala, como a exclusão dos itens Câimbra e Vertigem, bem como o agrupamento de categorias de resposta para alguns itens (11). Não se tem conhecimento de estudos avaliando a SARA sob a perspectiva da TRI. Em geral, as escalas NESSCA e SARA são correlacionadas mas, apesar de ambas somarem os mesmos 40 pontos, não se sabe como comparar cada pontuação individualmente.

O objetivo principal deste artigo é apresentar o método de equalização de escalas e explorar sua utilização com a abordagem da TRI no contexto da saúde. Para ilustrar a utilização do método será construída uma relação entre os instrumentos de medida para avaliação do nível de comprometimento pela doença de Machado-Joseph, SARA e NESSCA, estabelecendo escores equivalentes entre as escalas.

## MÉTODOS

### Fontes de Dados[*]

Diversas fontes de dados foram utilizadas, referentes a estudos procedidos com pacientes diagnosticados com SCA3/MJD do Hospital de Clínicas de Porto Alegre (HCPA):

a) Avaliações da NESSCA de 156 pacientes de um estudo de história natural descrito por Jardim et al. (12);

---

[*] Os estudos que originaram os dados para este trabalho foram aprovados pelo comitê de ética do Hospital de Clínicas de Porto Alegre: (a) GPPG-HCPA-02194; (b) GPPG-HCPA-09418; (c) GPPG-HCPA-13-0303; (d) GPPG-HCPA-06-0613; (e) GPPG-HCPA-14-0625.

b) Avaliações da NESSCA e SARA basais de 60 pacientes de um ensaio clínico randomizado de Saute et al. (13);

c) Avaliações de NESSCA e SARA de 35 pacientes de um estudo descrito por Oliveira et al. (14);

d) Avaliações de NESSCA e SARA de 24 pacientes do estudo de Saute et al. (15);

e) Avaliações de NESSCA e SARA de 8 pacientes com início na infância descritos por Donis et al. (16).

Os dados foram divididos em dois grupos, o Grupo 1 compreende os pacientes que possuem apenas avaliações da NESSCA, estudo (a), e o Grupo 2 é composto pelos pacientes restantes, estudos (b), (c), (d), e (e), que possuem avaliações para NESSCA e SARA. No caso de estudos prospectivos, com mais de uma medida das escalas (estudos (a), (b), (c) e (e)), foi utilizada a mais antiga dentre as que estavam completas, sendo descartadas as restantes. Pacientes com ausência de informação para algum item foram excluídos da análise, sendo 53 da fonte de dados (a) e um da fonte (e). Dada a importância do sintoma de marcha para a doença, comum às escalas NESSCA e SARA, os únicos dois pacientes que apresentaram ausência desse sintoma em pelo menos uma das duas escalas foram excluídos, sendo um da fonte (c) e um da fonte (e). Por fim, a amostra foi composta por 103 pacientes do Grupo 1 e 124 do Grupo 2. Outras informações como idade, gênero, duração da doença no momento da avaliação e número de repetições do CAG expandido também foram coletadas.

**Delineamento de Equalização**

A equalização pode ser feita a partir de grupos equivalentes ou de grupos não equivalentes. Os delineamentos de grupos equivalentes (em inglês, *Random Groups* - RG e *Single Groups* - SG) partem do princípio de que todos os indivíduos respondentes pertencem a uma mesma população. Por outro lado, se essa proposição não for atendida, ainda assim é possível lidar com grupos não equivalentes (em inglês, *Common Item Nonequivalent Groups* - CINEG), desde que existam itens comuns entre os instrumentos de medida, os quais irão servir para fazer uma conexão entre as diferentes populações (17). No contexto da SCA3/MJD, o delineamento mais adequado para a equalização das escalas foi o de grupos não equivalentes com itens comuns (CINEG). Os itens comuns devem ser representativos e não há uma regra estabelecida a respeito da quantidade, mas alguns autores recomendam que 40% do total de itens sejam comuns entre os testes (18). Além disso, foram utilizados apenas os itens da NESSCA cerebelar, ou seja, foram descartados os itens que avaliam sintomas não cerebelares da NESSCA, de forma a garantir a unidimensionalidade das escalas (11). A SARA e a NESSCA cerebelar compartilham de dois itens muito semelhantes que avaliam os sintomas Marcha (SARA)/Ataxia de Marcha (NESSCA) e Coordenação da Fala (SARA)/Disartria (NESSCA). A quantidade de categorias foi adaptada e encontra-se detalhada no Anexo E, Quadros 1 e 2. Com estas alterações, o escore máximo da SARA foi reduzido para 34 pontos ao invés dos 40 pontos originais (redução de quatro pontos no item Marcha e dois pontos no item Coordenação da Fala). Além disso, outro ajuste feito na SARA foi nos itens que avaliam os lados direito e esquerdo do corpo, separadamente, permitindo pontuações não inteiras para a média dos dois lados. Os valores 0,5; 1,5; 2,5 e 3,5 foram arredondados para 1; 2; 3 e 4, respectivamente, do contrário, na etapa computacional, tais pontuações não inteiras configurariam mais categorias de resposta. A NESSCA cerebelar passou a somar de 0 a 15 pontos devido a exclusão dos itens não cerebelares.

**Análise Estatística**

A primeira etapa da análise estatística teve por objetivo a calibração dos parâmetros dos itens, para isso, existem três métodos de calibração que podem ser empregados no delineamento CINEG: calibração dos parâmetros fixos, simultânea ou separada. Para este exercício foi escolhida a calibração separada por ser o método mais utilizado e que abrange mais etapas do processo (17,18). Este método consiste em calibrar separadamente os dados de cada instrumento de medida, ou seja, os parâmetros dos itens da NESSCA cerebelar foram estimados separadamente dos parâmetros dos itens da SARA. Para isso, foram utilizadas as avaliações da NESSCA cerebelar do Grupo 1 e as avaliações da SARA do Grupo 2 – descartando momentaneamente as avaliações da NESSCA cerebelar do Grupo 2. Os modelos TRI adequados para modelar os dados das escalas são o GRM (em inglês, *Graded Response Model*) e o GPCM (em inglês, *Generalized Partial Credit Model*), pois são modelos apropriados para dados politômicos possuem categorias ordenadas, não necessariamente na mesma quantidade. As fórmulas dos modelos encontram-se no Anexo A do artigo. Para a equalização as duas escalas devem ser ajustadas pelo mesmo modelo, o mais adequado foi escolhido através dos critérios de seleção AIC e BIC.

A utilização do GRM e do GPCM requer que duas suposições sejam satisfeitas: a independência local (tomando como base o nível do traço latente, os itens não devem ser correlacionados uns com os outros) e unidimensionalidade (somente um traço latente está sendo avaliado). Se a suposição de unidimensionalidade estiver atendida, então a independência local também estará satisfeita (19,20). A unidimensionalidade foi avaliada através da análise fatorial – autores sugerem que o primeiro fator deve explicar no mínimo 20% da variância total (21). A contribuição dos itens foi avaliada observando as estimativas dos parâmetros, a Curva Característica do Item (CCI), que descreve a relação ente a probabilidade e o traço latente, e a Curva de Informação do Item (CII), que permite avaliar a discriminação dos itens em modelos politômicos (5).

O resultado da calibração separada são dois traços latentes estimados e parâmetros dependentes dos grupos e, por isso, não estão na mesma métrica. Logo, é necessária uma etapa de transformação linear, a qual permite criar uma relação entre as escalas através dos parâmetros dos itens comuns (17,18). Conforme definições encontradas em Kolen e Brennan (17), pela propriedade de invariância dos parâmetros dos modelos TRI, existe uma relação linear de forma que:

$$\theta_S = A\theta_N + B, \tag{1}$$

onde $A$ é o coeficiente angular, $B$ é o intercepto e $\theta_S$ e $\theta_N$ são os traços latentes estimados para cada um dos instrumentos de medida, SARA e NESSCA cerebelar, respectivamente.

Para os modelos GRM e GPCM as estimativas dos parâmetros dos itens comuns se relacionam da seguinte forma (18):

$$a_{iS} = a_{iN}/A \tag{2}$$

e

$$b_{iS} = Ab_{iN} + B \tag{3}$$

e

$$c_{iS} = c_{iN} \tag{4}$$

onde $a_{iS}$, $b_{iS}$ e $c_{iS}$ são os parâmetros dos itens para o item $i$ na escala SARA e $a_{iN}$, $b_{iN}$ e $c_{iN}$ são os parâmetros dos itens para o item $i$ na escala NESSCA cerebelar.

Existem diferentes métodos de estimação das constantes $A$ e $B$, os principais são os métodos dos momentos (*Mean/Mean* e *Mean/Sigma*), conhecidos pela simplicidade, e os métodos de curva característica (*Haebara* e *Stocking-Lord*), que são empregados quando se busca por mais robustez (22). Existem estudos com dados simulados comparando a eficiência dos quatro métodos, onde os métodos da curva característica obtiveram, no geral, melhor desempenho, embora os autores ressaltem que os resultados podem não ser os mesmos para dados reais (22,23). Assim, os quatro métodos foram apurados para comparação das estimativas $A$ e $B$ que relacionam $\theta_S$ e $\theta_N$.

No entanto, o resultado em termos do traço latente pode ser de difícil compreensão, geralmente o público está habituado a valores de escore absolutos, como na TCT (17,18). Em função disto, foram desenvolvidas duas técnicas que podem relacionar, de fato, um escore absoluto na NESSCA a um escore absoluto na SARA, mesmo que tenham sido avaliadas sob a perspectiva da TRI, facilitando a interpretação. Estes métodos são conhecidos por equalização do verdadeiro escore (em inglês, *True Score Equating* - TSE) e equalização do escore observado (em inglês, *Observed Score Equating* – OSE), cujo resultado é uma relação direta que associa um escore na SARA a um escore equivalente na NESSCA, em termos de $\theta$ (18). Autores obtiveram resultados semelhantes para os métodos OSE e TSE quando utilizado o delineamento CINEG (24), embora o OSE possa levar a resultados fora dos limites da escala. Neste exercício foi abordado o método TSE.

Em um segundo momento, o objetivo foi avaliar o resultado do procedimento do TSE sob a perspectiva dos diferentes métodos de transformação linear. Os indivíduos do Grupo 2, com medidas reais para as duas escalas, possibilitaram essa avaliação. Utilizando como base o escore SARA observado e, através da tabela resultante da TSE, foi encontrado o escore NESSCA equivalente, viabilizando a comparação com o escore NESSCA observado. Os métodos de transformação linear foram avaliados utilizando medidas descritivas (média, quartis, mínimo e máximo), bem como a análise gráfica de *Bland-Altman*.

### Aspectos Computacionais

As análises foram realizadas no software R versão 3.4.4 (25). A análise dos fatores foi feita utilizando o pacote *nFactors* (26), os parâmetros foram calibrados com funções do pacote *ltm* (27) e a transformação linear e a equalização do verdadeiro escore foram feitas utilizando o pacote *plink* (28). Os gráficos de *Bland-Altman* foram feitos com o auxílio do pacote *blandr* (29). No Anexo D do artigo constam algumas das etapas mais relevantes da sintaxe.

## RESULTADOS

Características dos pacientes estão apresentadas na Tabela 1. Os grupos são homogêneos em termos das características clínicas inerentes a doença. Após as modificações da etapa descrita nos métodos, o intervalo de escore possível para a NESSCA cerebelar passou a ser (0,15) e, para a SARA ajustada, (0,34).

O resultado da análise fatorial sugere que a suposição de unidimensionalidade está atendida (21) por que há um fator preponderante que explica 59,6% e 48,1% da variância total para os dados da SARA e NESSCA cerebelar, respectivamente.

**Tabela 1.** Características da amostra.

| Características | Grupo 1 n = 103 | Grupo 2 n = 124 |
|---|---|---|
| **Gênero*** | | |
| Feminino | 51 (49,5) | 69 (55,6) |
| Masculino | 51 (49,5) | 55 (44,4) |
| Dados faltantes | 1 (1,0) | |
| **Características clínicas[†]** | | |
| Idade no início da doença, em anos | 34,4 (9,9) (7 - 57) | 33,4 (11,8) (5 - 65) |
| Duração da doença no momento da avaliação, em anos | 9,5 (5,7) (0 - 29) | 8,3 (4,7) (1 - 26) |
| CAG expandido, em número de repetições | 74,2 (2,4) (69 - 79) | 75,3 (3,6) (68 - 91) |
| **Escalas de medida[†]** | | |
| NESSCA cerebelar | 7,4 (2,7) (1 - 14) | 7,0 (2,4) (2 - 15) |
| SARA ajustada | - | 12,8 (6,1) (3 - 34) |

**\* Avaliados em valores absolutos (percentual).**

**[†] Avaliados em média (desvio padrão) (mínimo - máximo).**

Foi escolhido o modelo GPCM para ajuste da NESSCA cerebelar e da SARA, em virtude dos resultados que minimizam os valores de AIC e BIC – o resultado detalhado está no Anexo B do artigo. A Tabela 2 apresenta o resultado da calibração dos parâmetros dos itens da escala NESSCA cerebelar pelo GPCM - as CCIs e as CIIs encontram-se no Anexo E do artigo, Figuras 1 a 10. Avaliando as estimativas dos parâmetros, em conjunto com as CIIs, podem ser considerados itens com maior poder de discriminação: Ataxia de Marcha, Ataxia nos Membros, Disartria e Disfagia. A Tabela 3, por sua vez, apresenta as estimativas dos parâmetros para os itens da SARA ajustada pelo modelo GPCM – as CCIs e as CIIs estão no Anexo F do artigo, Figuras 11 a 27. No geral, todos os itens da SARA apresentaram bom poder de discriminação. Estão destacados na cor cinza, nas Tabelas 2 e 3, os itens comuns entre as escalas. Nessa etapa, ainda que as estimativas dos parâmetros sejam semelhantes para os itens comuns, os mesmos ainda são dependentes da amostra de respondentes.

**Tabela 2.** Estimativas para os parâmetros dos itens da NESSCA cerebelar.

| Item | | $a$ | $b_1$ | $b_2$ | $b_3$ | $b_4$ |
|---|---|---|---|---|---|---|
| 1 | Ataxia de Marcha | 4,530 | -1,567 | 0,192 | 1,131 | |
| 2 | Ataxia nos Membros | 0,741 | -2,533 | -0,534 | 1,036 | |
| 3 | Nistagmo | 0,529 | -1,841 | 4,960 | | |
| 4 | Disartria | 1,796 | -1,427 | 0,889 | 1,802 | 1,779 |
| 5 | Disfagia | 1,623 | -0,955 | 0,857 | | |

**Tabela 3.** Estimativas para os parâmetros dos itens da SARA ajustada.

| Item | | $a$ | $b_1$ | $b_2$ | $b_3$ | $b_4$ | $b_5$ | $b_6$ |
|---|---|---|---|---|---|---|---|---|
| 1 | Marcha | 4,976 | -1,259 | 0,623 | 2,155 | | | |
| 2 | Equilíbrio de Pé | 1,750 | -1,406 | -1,049 | 0,833 | 1,754 | 0,837 | 1,784 |
| 3 | Equilíbrio Sentado | 1,959 | 0,801 | 1,634 | 2,084 | 2,437 | | |
| 4 | Coordenação da Fala | 1,477 | -1,765 | 0,699 | 1,769 | 2,395 | | |
| 5 | Teste de Perseguição do Dedo | 1,246 | -2,893 | -0,063 | 1,815 | 2,246 | | |
| 6 | Teste Dedo-Nariz | 1,244 | -0,827 | 0,903 | 2,292 | 2,136 | | |
| 7 | Diadococinesia | 0,983 | -1,217 | 0,454 | -0,282 | 3,225 | | |
| 8 | Teste Calcanhar-Joelho-Canela | 1,049 | -2,868 | -0,107 | 0,461 | 1,924 | | |

Após a calibração dos parâmetros dos itens, foi feita a transformação linear para que os traços latentes, $\theta_N$ e $\theta_S$, fiquem na mesma escala. Os resultados com as estimativas para as constantes $A$ e $B$ dos quatro métodos avaliados – *Mean/Mean*, *Mean/Sigma*, *Haebara* e *Stocking-Lord* – estão no Anexo G do artigo e foram estimados com base nas equações (4) e (5). A partir deste resultado, foi calculada a relação de verdadeiro escore sob a perspectiva dos quatro métodos. Para fins de ilustração, será utilizado o resultado do método *Mean/Sigma*, o qual está apresentado na Tabela 4 para duas situações: partindo do escore SARA para obter o escore NESSCA equivalente e partindo do escore NESSCA para obter o escore SARA equivalente - a relação é simétrica. Os valores de $\theta_N$ e $\theta_S$ se relacionam conforme equação (3) utilizando as constantes $A$ e $B$ estimadas.

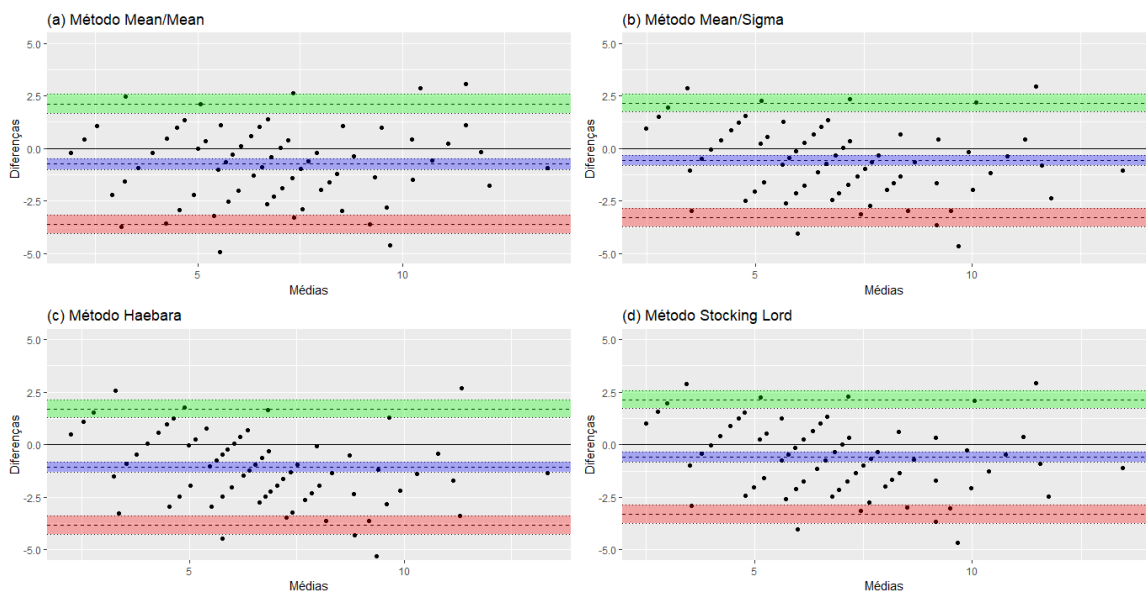

**Figura 1.** Gráficos *Bland-Altman* das diferenças entre o escore NESSCA estimado pela SARA e o escore NESSCA observado: (a) Método *Mean/Mean*; (b) Método *Mean/Sigma*; (c) Método *Haebara*; (d) Método *Stocking-Lord*.

**Tabela 4.** Equalização do verdadeiro escore utilizando o método *Mean/Sigma*[†]

| TSE: SARA para NESSCA | | |
|---|---|---|
| **Escore SARA observado** | **Escore NESSCA estimado pela SARA** | $\theta_S$ |
| 0 | 0,000 | -61,37 |
| 1 | 0,829 | -3,00 |
| 2 | 1,332 | -2,30 |
| 3 | 2,097 | -1,81 |
| 4 | 3,053 | -1,49 |
| 5 | 3,595 | -1,29 |
| 6 | 3,976 | -1,11 |
| 7 | 4,384 | -0,88 |
| 8 | 4,812 | -0,60 |
| 9 | 5,177 | -0,33 |
| 10 | 5,484 | -0,09 |
| 11 | 5,813 | 0,13 |
| 12 | 6,242 | 0,32 |
| 13 | 6,704 | 0,48 |
| 14 | 7,081 | 0,62 |
| 15 | 7,395 | 0,75 |
| 16 | 7,696 | 0,89 |
| 17 | 8,003 | 1,03 |
| 18 | 8,310 | 1,15 |
| 19 | 8,631 | 1,27 |
| 20 | 8,993 | 1,38 |
| 21 | 9,410 | 1,49 |
| 22 | 9,858 | 1,61 |
| 23 | 10,272 | 1,74 |
| 24 | 10,617 | 1,86 |
| 25 | 10,899 | 1,97 |
| 26 | 11,140 | 2,08 |
| 27 | 11,360 | 2,18 |
| 28 | 11,582 | 2,29 |
| 29 | 11,827 | 2,42 |
| 30 | 12,107 | 2,59 |
| 31 | 12,429 | 2,82 |
| 32 | 12,794 | 3,24 |

[†] Devido a pequena quantidade ou a ausência de pacientes para os escores mais altos, só foi possível estimar a relação até o escore 32 para a SARA e até o escore 13 para a NESSCA.

Com esta relação de "de-para" estabelecida na Tabela 4 e as avaliações da NESSCA do Grupo 2 (as quais não foram utilizadas na etapa de calibração), foi possível avaliar a acurácia de cada um dos métodos. Utilizando o resultado presente nas duas primeiras colunas da Tabela 4 e os dados dos pacientes do Grupo 2, foi calculada a diferença entre o escore NESSCA estimado pela SARA via equalização, e o escore NESSCA observado para o método *Mean/Sigma*. O mesmo foi realizado para os demais métodos. A Figura 1 ilustra, através dos gráficos de *Bland-Altman*, a diferença entre o escore NESSCA estimado pela SARA e o escore NESSCA observado, para os quatro métodos de transformação linear.

Na Tabela 5 estão as medidas descritivas das diferenças apresentadas entre o escore NESSCA estimado pela SARA e o escore NESSCA observado para a amostra de 124 pacientes do Grupo 2. Pela análise gráfica e descritiva, os métodos *Mean/Sigma* e *Stocking-Lord* apresentaram melhores resultados, a mediana das diferenças foi de 0,78 e 0,81, respectivamente. O método *Mean/Sigma* foi escolhido para ilustrar a relação de equivalência entre as escalas, conforme Tabela 4, por ter média e mediana das diferenças mais próximas de zero.

**Tabela 5.** Medidas descritivas das diferenças entre o escore NESSCA estimado pela SARA e o escore NESSCA observado.

| Método de transformação | Mínimo | Q1 | Mediana | Média | Q3 | Máximo |
|---|---|---|---|---|---|---|
| *Mean/Mean* | 0,061 | 0,401 | 0,939 | 1,240 | 1,783 | 4,779 |
| *Mean/Sigma* | 0,023 | 0,366 | 0,823 | 1,144 | 1,740 | 4,604 |
| *Haebara* | 0,022 | 0,581 | 1,248 | 1,465 | 2,116 | 5,392 |
| *Stocking-Lord* | 0,006 | 0,378 | 0,869 | 1,173 | 1,828 | 4,679 |

## DISCUSSÃO

A equalização de escalas, no contexto da área da saúde, é uma técnica com muito potencial e ainda pouco explorada (2,3). Através da equalização é possível comparar resultados de publicações científicas que utilizam diferentes escalas de avaliação, para qualquer segmento da saúde que faça uso de escalas de medida. Neste trabalho, foi possível explorar os conceitos e as etapas da equalização de escalas trazendo para o contexto da área da saúde ao utilizar, de forma ilustrativa, dados de pacientes portadores da doença de Machado-Joseph. Como resultado da aplicação da técnica, foi possível estabelecer uma sugestão de relação direta entre a NESSCA cerebelar e a SARA. Com o método de transformação linear *Mean/Sigma* metade das estimativas tiveram erro abaixo de 0,82 pontos no escore NESSCA cerebelar estimado pela SARA em comparação com o escore NESSCA observado nos pacientes. Além disso, no nosso conhecimento, a avaliação da SARA pela perspectiva da TRI é inédita, e nessa análise as estimativas para os parâmetros de todos os itens mostraram-se, no geral, bastante consistentes e os itens são representativos do comprometimento pela doença, corroborando com a consistência do instrumento (9).

O delineamento de grupos não equivalentes, CINEG, é um dos mais utilizados na prática (17,18). Isso ocorre devido a flexibilidade e possível redução nos custos de aplicação, pois nem sempre é possível submeter os pacientes a todas as escalas, como acontece em outros delineamentos (17,18). Para estabelecer uma relação de equalização, é recomendado que os testes sejam construídos com este propósito, e no caso do CINEG, atendendo a requisitos a respeito da estrutura dos itens comuns, como a ordem em que os itens aparecem e a semelhança dos enunciados – tais recomendações evitam a presença de erros sistemáticos (17). Outro erro que pode estar presente na equalização é erro aleatório, este pode ser reduzido a medida que se aumenta o tamanho da amostra, até que se torne desprezível. Amostras pequenas, abaixo de 100 indivíduos, podem levar a erros altos (17). Além disso, a quantidade de itens comuns deve ser representativa em relação ao total de itens do instrumento, não há uma regra estabelecida, porém alguns autores sugerem que 40% do total de itens sejam comuns entre os testes (18).

Na área da saúde, satisfazer a todas estas recomendações pode ser uma tarefa bastante desafiadora, visto que na maioria dos casos os instrumentos de medida foram construídos sem o propósito de equalizar, e, consequentemente, não possuem itens idênticos e que aparecem na mesma ordem em ambos os testes. A técnica exige uma série de cuidados e é fácil entender porque sua aplicação é frequente na avaliação educacional, com literatura majoritariamente oriunda das ciências sociais e comportamentais, e tão incomum na área da saúde (17,18). No entanto, aqui é importante ressaltar que, na educação, por exemplo, a prova anual do SAEB (Sistema de Avaliação da Educação Básica), tem por finalidade realizar um diagnóstico da educação básica brasileira, buscando avaliar os alunos em uma mesma escala de conhecimento, mesmo que sejam submetidos a diferentes cadernos de prova (1). Nesse caso, a equalização tem papel fundamental na avaliação dos alunos, por isso é extremamente importante que todas as condições de construção e aplicação dos testes sejam atendidas, com o objetivo de obter uma equalização bem-sucedida. Enquanto que, na área da saúde, a possibilidade de comparar indivíduos aferidos por diferentes escalas pode ser um fator complementar na avaliação do paciente, que, no caso da SCA3/MJD, poderá auxiliar na administração dos tratamentos para contornar as manifestações clínicas com base em publicações científicas, proporcionando melhor qualidade de vida aos pacientes.

Existem limitações nos resultados da equalização, visto que a SARA e a NESSCA não foram construídas sob a mesma perspectiva e não possuíam itens idênticos, por isso foram necessárias adaptações nas categorias, de forma a forçar a presença de itens comuns. Ainda que, para a NESSCA cerebelar, a recomendação de itens comuns tenha sido atingida (40% de itens comuns), no caso da SARA, apenas 25% dos itens eram comuns, o que constitui uma possível limitação deste resultado. Para estudos futuros, uma alternativa para aumentar a quantidade de itens comuns seria relacionar o item Ataxia nos Membros da NESSCA cerebelar com uma combinação dos itens Teste de Perseguição do Dedo, Teste Dedo-nariz e Diadococinesia da SARA.

Vale lembrar que, na prática, dois pacientes com o mesmo escore SARA podem ter escores NESSCA diferentes, isso ocorre devido à variabilidade fenotípica da doença e aos itens não comuns. A relação obtida só seria perfeita em todos os casos se os instrumentos de medida equalizados fossem idênticos (17,18). Uma das propriedades da equalização, em inglês, *first-order equity property*, esclarece que é esperado que os pacientes obtivessem o mesmo escore equalizado caso fossem submetidos a outra escala, em média, mas não exige que as distribuições de probabilidade condicionais dos escores sejam iguais entre as escalas (18). Assim, é importante enxergar a equalização como uma sugestão de relação entre as escalas de medida, que viabilize a comparação de resultados científicos.

Este artigo teve por objetivo ilustrar o potencial da equalização de escalas para a área da saúde e demonstrar seu uso, através dos dados de pacientes portadores da doença de Machado-Joseph avaliados por duas escalas diferentes, NESSCA e SARA. Estudos futuros podem ampliar o uso da técnica para lidar com outras doenças e desfechos clínicos avaliados por diferentes escalas de medida.

A ser enviado a Revista Brasileira de Epidemiologia.

## REFERÊNCIAS

1. de Andrade DF, Tavares HR, Valle R da C. Teoria da Resposta ao Item: Conceitos e Aplicações. São Paulo, SP: ABE; 2000. 154 p.

2. Chen W-H, Revicki DA, Lai J-S, Cook KF, Amtmann D. Linking pain items from two studies onto a common scale using item response theory. J Pain Symptom Manage. 2009;38(4):615–28.

3. McHorney CA, Cohen AS. Equating health status measures with item response theory: illustrations with functional status items. Med Care. 2000;38(9 Suppl):II43-59.

4. Coluci MZO, Alexandre NMC, Milani D. Construção de instrumentos de medida na área da saúde. Ciência & Saúde Coletiva. 2015;20(3):925–36.

5. Castro SMJ, Trentini C, Riboldi J. Item response theory applied to the Beck Depression Inventory. Revista Brasileira de Epidemiologia. 2010;13(3):487–501.

6. Moreira Jr. FJ. Aplicações da teoria da resposta ao item (TRI) no Brasil. Revista Brasileira de Biometria. 2010;28(4):137–70.

7. Kieling C, Prestes PR, Saraiva-Pereira ML, Jardim LB. Survival estimates for patients with Machado–Joseph disease (SCA3). Clinical genetics. 2007;72(6):543–545.

8. Saute JAM, Jardim LB. Machado Joseph disease: clinical and genetic aspects, and current treatment. Expert Opinion on Orphan Drugs. 2015;3(5):517–535.

9. Schmitz-Hübsch T, du Montcel ST, Baliko L, Berciano J, Boesch S, Depondt C, et al. Scale for the assessment and rating of ataxia: development of a new clinical scale. Neurology. 2006 Jun 13;66(11):1717–20.

10. Kieling C, Rieder CRM, Silva ACF, Saute JAM, Cecchin CR, Monte TL, et al. A neurological examination score for the assessment of spinocerebellar ataxia 3 (SCA3). European journal of neurology. 2008;15(4):371–376.

11. Maciel TH. Aplicação da Teoria da Resposta ao Item ao escore NESSCA de avaliação da progressão da Doença de Machado Joseph. [Porto Alegre]: Universidade Federal do Rio Grande do Sul; 2013.

12. Jardim LB, Hauser L, Kieling C, Saute JAM, Xavier R, Rieder CRM, et al. Progression Rate of Neurological Deficits in a 10-Year Cohort of SCA3 Patients. The Cerebellum. 2010;9(3):419–28.

13. Saute JAM, de Castilhos RM, Monte TL, Schumacher-Schuh AF, Donis KC, D'Ávila R, et al. A randomized, phase 2 clinical trial of lithium carbonate in Machado-Joseph disease: Lithium Trial in Machado-Joseph Disease. Movement Disorders. 2014;29(4):568–73.

14. Oliveira CM, Reckziegel ER, Augustin MC, Rocha AG, Bolzan G, Santos JA, et al. Causal factors behind early- and late-onset Machado-Joseph disease patients do not interfere with the rate of neurological deterioration. In Pisa; 2017. Available from: http://www.iarc2017.com/wp-content/uploads/2017/09/IARC-Abstract-Book.pdf

15. Saute JAM, da Silva ACF, Souza GN, Russo AD, Donis KC, Vedolin L, et al. Body Mass Index is Inversely Correlated with the Expanded CAG Repeat Length in SCA3/MJD Patients. The Cerebellum. 2012;11(3):771–4.

16. Donis KC, Saute JAM, Krum-Santos AC, Furtado GV, Mattos EP, Saraiva-Pereira ML, et al. Spinocerebellar ataxia type 3/Machado-Joseph disease starting before adolescence. neurogenetics. 2016;17(2):107–13.

17. Kolen MJ, Brennan RL. Test equating, scaling, and linking: methods and practices. Third Edition. New York: Springer; 2014. 566 p. (Statistics for Social and Behavioral Sciences).

18. Nering ML, Ostini R. Handbook of polytomous item response theory models. New York, NY: Routledge; 2010. 296 p.

19. Hambleton RK, Swaminathan H, Rogers HJ. Fundamentals of item response theory. Newbury Park, Calif: Sage Publications; 1991. 174 p. (Measurement methods for the social sciences series).

20. Hays RD, Morales LS, Reise SP. Item response theory and health outcomes measurement in the 21st century. Med Care. 2000;38(9 Suppl):II28-42.

21. Hattie J. Methodology Review: Assessing Unidimensionality of Tests and ltems. Applied Psychological Measurement. 1985;9(2):139–64.

22. Kim S, Lee W-C. An Extension of Four IRT Linking Methods for Mixed-Format Tests. Journal of Educational Measurement. 2006;43(1):53–76.

23. Hanson BA, Béguin AA. Obtaining a Common Scale for Item Response Theory Item Parameters Using Separate Versus Concurrent Estimation in the Common-Item Equating Design. Applied Psychological Measurement. 2002;26(1):3–24.

24. Lord FM, Wingersky MS. Comparison of IRT True-Score and Equipercentile Observed-Score "Equatings." Applied Psychological Measurement. 1984;8(4):453–61.

25. R Core Team. R: A Language and Environment for Statistical Computing [Internet]. Vienna, Austria: R Foundation for Statistical Computing; 2018. Available from: https://www.R-project.org/

26. Raiche G. **nFactors**: An *R* package for parallel analysis and non graphical solutions to the Cattell scree test. 2010; Available from: http://CRAN.R-project.org/package=nFactors

27. Rizopoulos D. **ltm** : An R package for Latent Variable Modelling and Item Response Theory Analyses. Journal of Statistical Software. 2006;17(5):1–25.

28. Weeks JP. **plink**: An R package for Linking Mixed-Format Tests Using IRT-Based Methods. Journal of Statistical Software. 2010;35(12):1–33.

29. Datta D. **blandr**: A Bland-Altman Method Comparison package for *R*. 2017; Available from: https://github.com/deepankardatta/blandr

# 9. ANEXOS DO ARTIGO

**ANEXO A**

**Modelos GRM e GPCM**

**Modelo de Resposta Gradual**

O Modelo de Resposta Gradual (Samejima, 1969), diferentemente do NRM, assume que as categorias de um item tenham uma ordem. Conhecido por GRM (em inglês, *Graded-Response Model*), trata-se de uma generalização do 2PL e é considerado um modelo TRI "indireto" pois requer um procedimento adicional para calcular a probabilidade condicional de um indivíduo ter um determinado nível do sintoma. Uma vantagem do GRM é que os itens do instrumento não precisam ter a mesma quantidade de categorias de resposta, como ocorre na NESSCA e na SARA. Considerando que as possíveis categorias de um item sejam denotadas por $k = 0,1, \ldots, m_i$ onde $m_i + 1$ é o número de categorias do item $i$, a probabilidade de um indivíduo $j$ pertencer a uma particular categoria ou outra mais alta pode ser dada por:

$$P_{i,k}^+(\theta_j) = \frac{1}{1 + e^{-Da_i(\theta_j - b_{i,k})}}, \tag{1}$$

com $i = 1,2, \ldots, I, j = 1,2, \ldots, n$ e $k = 0,1, \ldots, m_i$, onde:

| | |
|---|---|
| $\theta_j$ | representa a intensidade do comprometimento pela SCA3/MJD (traço latente) do $j$-ésimo paciente; |
| $a_i$ | é o parâmetro de inclinação comum a todas as categorias de um mesmo item $i$; |
| $b_{i,k}$ | é o parâmetro de posição da $k$-ésima categoria do item $i$, ou seja, representa o nível de comprometimento necessário para a escolha da categoria de resposta $k$, ou acima de $k$, com probabilidade igual a 0,50; |
| $D$ | é um fator de escala, constante e igual a 1. Utiliza-se 1,7 quando se deseja que a função logística forneça resultados semelhantes ao da função ogiva normal; |
| $I$ | é o número de itens no instrumento de medida; |
| $n$ | é o número de indivíduos respondentes. |

Deverá existir uma ordenação entre as categorias de um dado item, ou seja:

$$b_{i,1} \leq b_{i,2} \leq \cdots \leq b_{i,m_i}$$

A probabilidade de um indivíduo $j$ pertencer a categoria $k$ no item $i$ é dada pela expressão:

$$P_{i,k}(\theta_j) = P_{i,k}^+(\theta_j) - P_{i,k+1}^+(\theta_j) \tag{2}$$

onde $P_{i,0}^+(\theta_j) = 1$ e $P_{i,m_i+1}^+(\theta_j) = 0$, logo:

$$P_{i,k}(\theta_j) = \frac{1}{1+e^{-Da_i(\theta_j-b_{i,k})}} - \frac{1}{1+e^{-Da_i(\theta_j-b_{i,k+1})}} \qquad (3)$$

O número de parâmetros, por item, será dado pelo número de categorias $k$ do item $i$.

Por se tratar de modelos para itens politômicos, com mais de duas categorias, a equação (3) gera as curvas de categoria de resposta (CCR), as quais são simbolizadas por $P_{ik}(\theta)$. Estas curvas ilustram a relação entre a probabilidade de um indivíduo com comprometimento $\theta$ pertencer a categoria $k$ do sintoma do item $i$. O conjunto de todas as CCRs de um teste resulta na curva característica do teste (CCT) que representa a probabilidade de obtenção de um escore total em função de $\theta$, pode ser interpretado como a média para um dado valor de $\theta$. Para um teste com $I$ itens, a CCT é dada por:

$$T(\theta) = \sum_{i=1}^{I} \sum_{k=1}^{m_i} U_{ik} P_{ik}(\theta) \qquad (4)$$

onde $U_{ik}$ é uma função do escore do item, ou seja, são os valores de escore possíveis de obter no item $i$.

Geralmente utiliza-se $U_{ik} = k - 1$ (quando uma resposta associada a primeira categoria recebe escore zero, que é o caso da NESSCA e da SARA) ou $U_{ik} = k$ (quando uma resposta associada a primeira categoria representa escore igual a 1) (Kolen e Brennan, 2014).

**Modelo de Crédito Parcial Generalizado**

O modelo de crédito parcial generalizado foi desenvolvido por Muraki, em 1992, e consiste de uma generalização do PCM, relaxando a hipótese de poder de discriminação igual para todos os itens (Muraki, 1992). Ou seja, permite que os itens dentro de uma escala tenham diferentes parâmetros de inclinação, o que é interessante no contexto da NESSCA e da SARA. Também conhecido como *Generalized Partial Credit Model* (GPCM), supondo que o item $i$ possui $m_i + 1$ categorias de resposta ordenadas ($k = 0, 1, \ldots, m_i$), temos que o modelo é dado por:

$$P_{i,k}(\theta_j) = \frac{exp\left[\sum_{u=0}^{k} Da_i(\theta_j - b_{i,u})\right]}{\sum_{u=0}^{m_i} exp\left[\sum_{v=0}^{u} Da_i(\theta_j - b_{i,v})\right]}, \qquad (5)$$

com $i = 1,2,\ldots,I, j = 1,2,\ldots,n$ e $k = 0,1,\ldots,m_i$, onde:

$\theta_j$ \qquad representa a intensidade do comprometimento pela SCA3/MJD (traço latente) do $j$-ésimo paciente;

$P_{i,k}(\theta_j)$ \qquad é a probabilidade de um indivíduo com nível de comprometimento $\theta_j$ ter um

| | sintoma na categoria $k$ dentre as $m_i + 1$ categorias do item $i$; |
|---|---|
| $a_i$ | é o parâmetro de inclinação do item $i$; |
| $b_{i,k}$ | é o parâmetro do item que regula a probabilidade do sintoma ser $k$ ao invés da categoria adjacente $(k-1)$ no item $i$. Cada parâmetro $b_{i,k}$ corresponde ao valor do traço latente no qual o indivíduo tem a mesma probabilidade de ser classificado nas categorias $k$ e $(k-1)$, isto é, onde $P_{i,k}(\theta_j) = P_{i,k-1}(\theta_j)$. Pode ser interpretado como um parâmetro de interseção entre as categorias de resposta do item $i$; |
| $D$ | é um fator de escala, constante e igual a 1. Utiliza-se 1,7 quando se deseja que a função logística forneça resultados semelhantes ao da função ogiva normal; |
| $I$ | é o número de itens no instrumento de medida; |
| $n$ | é o número de indivíduos respondentes. |

É importante observar que o parâmetro de inclinação ($a_i$) presente neste modelo não deve ser interpretado diretamente, da mesma forma como nos modelos dicotômicos. Nos modelos politômicos, a discriminação do item depende da combinação de $a_i$ com a distribuição dos parâmetros $b_{i,k}$. Os parâmetros $b_{i,k}$ são os pontos na escala onde as curvas das categorias se cruzam, em qualquer ponto da escala $\theta_j$. No geral, define-se $b_{i,0} = 0$. Além disso, frequentemente os parâmetros $b_{i,k}$ são decompostos em um parâmetro de posição $b_i$ e nos parâmetros para as categorias, $d_{i,k}$, onde:

$$b_{i,k} = b_i - d_{i,k} \tag{6}$$

Para avaliar a contribuição de um item politômico pode-se observar a Curva de Informação do Item (CII). Essa curva indica a quantidade de informação que um determinado sintoma contribui para a medida do traço latente e em qual intervalo esse sintoma é mais informativo (Castro et al., 2010).

**ANEXO B**

**Resultado do ajuste dos modelos**

**Tabela 1.** Resultado do ajuste dos modelos para a NESSCA cerebelar.

| Modelo | AIC | BIC |
|--------|---------|---------|
| GPCM | 1042,66 | 1092,72 |
| GRM | 1038,08 | 1088,14 |

**Tabela 2.** Resultado do ajuste dos modelos para a SARA.

| Modelo | AIC | BIC |
|--------|---------|---------|
| GPCM | 2106,07 | 2221,71 |
| GRM | 2114,17 | 2229,80 |

**ANEXO C**

**Adaptação das categorias da SARA**

**Quadro 3.** Adaptação das categorias da SARA para o item Marcha.

| NESSCA | | SARA | |
|---|---|---|---|
| **Gravidade** | **Categoria** | **Gravidade** | **Categoria** |
| Ausente. | 0 | Normal, sem dificuldade para andar, virar-se ou andar na posição pé-ante-pé (até um erro aceito). | 0 |
| Mínima: apenas ao andar na ponta dos pés, com os calcanhares e em conjunto. | 1 | Discretas dificuldades, somente visíveis quando anda 10 passos consecutivos na posição pé-ante-pé. | 1 |
| | | Claramente anormal, marcha na posição pé-ante-pé impossível com 10 ou mais passos. | 2 |
| Moderado: autonomia de marcha preservada. | 2 | Consideravelmente cambaleante dificuldades na meia-volta, mas ainda sem apoio. | 3 |
| | | Mareadamente cambaleante, necessitando de apoio intermitente da parede. | 4 |
| Incapacidade de caminhar sem ajuda. | 3 | Gravemente cambaleante, apoio permanente com uma bengala ou apoio leve de um braço. | 5 |
| | | Marcha > 10m somente possível com apoio forte (2 bengalas especiais ou um andador ou um acompanhante). | 6 |
| | | Macha < 10m somente possível com um apoio forte (2 bengalas especiais ou um andador ou um acompanhante). | 7 |
| Cadeira de rodas ou acamados. | 4 | Incapaz de andar mesmo com apoio. | 8 |

**Quadro 4.** Adaptação das categorias da SARA para o item Coordenação da Fala.

| NESSCA | | SARA | |
|---|---|---|---|
| **Gravidade** | **Categoria** | **Gravidade** | **Categoria** |
| Ausente. | 0 | Normal | 0 |
| Leve: Dificuldade de fala, mas fácil de entender. | 1 | Sugestivo de alteração na fala. | 1 |
| | | Alteração na fala, mas fácil de entender. | 2 |
| Moderado: discurso compreensível, mas com dificuldade. | 2 | Ocasionalmente palavras difíceis de entender. | 3 |
| Grave: discurso de difícil compreensão. | 3 | Muitas palavras difíceis de entender. | 4 |
| | | Somente palavras isoladas compreensíveis. | 5 |
| Anartria. | 4 | Fala ininteligível/anartria. | 6 |

# ANEXO D

## Sintaxe

Este anexo apresenta algumas partes da sintaxe consideradas mais relevantes para este trabalho.

Análise Fatorial

```
N<- cor(nessca[,6:10])
componentAxis(N, nFactors=3)
S<-cor(sara[6:13])
componentAxis(S, nFactors=3)
```

Calibração dos Parâmetros – NESSCA

```
t2_nessca <- gpcm(nessca[,6:10])
```

Calibração dos Parâmetros – SARA

```
t1_sara <- gpcm(sara[,6:13])
```

Curva Característica do Item (Exemplo: NESSCA)

```
plot(t2_nessca, type = "ICC", items=1, main="Curva Característica do Item\nAtaxia de Marcha",xlab="θ",ylab="P(θ)")
plot(t2_nessca, type = "ICC", items=2, main="Curva Característica do Item\nAtaxia nos Membros",xlab="θ",ylab="P(θ)")
plot(t2_nessca, type = "ICC", items=3, main="Curva Característica do Item\nNistagmo",xlab="θ",ylab="P(θ)")
plot(t2_nessca, type = "ICC", items=4, main="Curva Característica do Item\nDisartria",xlab="θ",ylab="P(θ)")
plot(t2_nessca, type = "ICC", items=5, main="Curva Característica do Item\nDisfagia",xlab="θ",ylab="P(θ)")
```

Curva de Informação do Item (Exemplo: NESSCA)

```
plot(t2_nessca, type = "IIC", items = 1, ylim = c(0,7), main="Curva de Informação do Item\nAtaxia de Marcha", xlab="θ",ylab="Informação")
plot(t2_nessca, type = "IIC", items = 2, ylim = c(0,7), main="Curva de Informação do Item\nAtaxia nos Membros", xlab="θ",ylab="Informação")
plot(t2_nessca, type = "IIC", items = 3, ylim = c(0,7), main="Curva de Informação do Item\nNistagmo", xlab="θ",ylab="Informação")
plot(t2_nessca, type = "IIC", items = 4, ylim = c(0,7), main="Curva de Informação do Item\nDisartria", xlab="θ",ylab="Informação")
plot(t2_nessca, type = "IIC", items = 5, ylim = c(0,7), main="Curva de Informação do Item\nDisfagia", xlab="θ",ylab="Informação")
```

Preparando input para equalização

Trata-se de uma lista que contém todas as informações necessárias para a equalização: parâmetros dos itens, quantidade de categorias de cada item, modelos TRI utilizados e itens comuns.

```
lista=list(pars=list(
        group1=data.frame(matrix(c(coef(t2_nessca)$Gait[[4]],coef(t2_nessca)$Limb[[4]],coef(t2_nessca)$Nistag[[3]],
coef(t2_nessca)$Disart[[5]],coef(t2_nessca)$Dysphag[[3]],coef(t2_nessca)$Gait[[1]],coef(t2_nessca)$Limb[[1]],coef(t2_
nessca)$Nistag[[1]],coef(t2_nessca)$Disart[[1]],coef(t2_nessca)$Dysphag[[1]],coef(t2_nessca)$Gait[[2]],coef(t2_nessca)
$Limb[[2]],coef(t2_nessca)$Nistag[[2]],coef(t2_nessca)$Disart[[2]],coef(t2_nessca)$Dysphag[[2]],coef(t2_nessca)$Gait
[[3]],coef(t2_nessca)$Limb[[3]],NA,coef(t2_nessca)$Disart[[3]],NA,NA,NA,NA,coef(t2_nessca)$Disart[[4]],NA),5,5)),
        group2=data.frame(matrix(c(coef(t1_sara)$Gait_A_cod[[4]],coef(t1_sara)$Stance[[7]],coef(t1_sara)$Sit[[5]],c
oef(t1_sara)$Speech_cod[[5]],coef(t1_sara)$Chase[[5]],coef(t1_sara)$Nosefing[[5]],coef(t1_sara)$Disdia[[5]],coef(t1_sa
ra)$HeelShin[[5]],coef(t1_sara)$Gait_A_cod[[1]],coef(t1_sara)$Stance[[1]],coef(t1_sara)$Sit[[1]],coef(t1_sara)$Speech
_cod[[1]],coef(t1_sara)$Chase[[1]],coef(t1_sara)$Nosefing[[1]],coef(t1_sara)$Disdia[[1]],coef(t1_sara)$HeelShin[[1]],
coef(t1_sara)$Gait_A_cod[[2]],coef(t1_sara)$Stance[[2]],coef(t1_sara)$Sit[[2]],coef(t1_sara)$Speech_cod[[2]],coef(t1_s
ara)$Chase[[2]],coef(t1_sara)$Nosefing[[2]],coef(t1_sara)$Disdia[[2]],coef(t1_sara)$HeelShin[[2]],coef(t1_sara)$Gait_
A_cod[[3]],coef(t1_sara)$Stance[[3]],coef(t1_sara)$Sit[[3]],coef(t1_sara)$Speech_cod[[3]],coef(t1_sara)$Chase[[3]],coe
f(t1_sara)$Nosefing[[3]],coef(t1_sara)$Disdia[[3]],coef(t1_sara)$HeelShin[[3]],NA,coef(t1_sara)$Stance[[4]],coef(t1_sa
ra)$Sit[[4]],coef(t1_sara)$Speech_cod[[4]],coef(t1_sara)$Chase[[4]],coef(t1_sara)$Nosefing[[4]],coef(t1_sara)$Disdia[[
4]],coef(t1_sara)$HeelShin[[4]],NA,coef(t1_sara)$Stance[[5]],NA,NA,NA,NA,NA,NA,coef(t1_sara)$Stance[[6]],N
A,NA,NA,NA,NA,NA),8,7))),
        cat = list(group1 = c(4,4,3,5,3), group2 = c(4,7,5,5,5,5,5,5)),
        items = list(group1 = list(grm = c(1:5)),
                     group2 = list(gpcm = c(1:8))),
        common = matrix(c(1,1,4,4),2,2,byrow = TRUE))
```

Definição de objetos auxiliares

```
pm1 <- as.poly.mod(5,"grm",lista$items$group1)
pm2 <- as.poly.mod(8,"gpcm",lista$items$group2)
x <- as.irt.pars(lista$pars,lista$common,lista$cat,list(pm1,pm2))
```

Equalização (Exemplo: NESSCA – Método *Mean/Mean*)

```
aux_mm_N <- plink(x, method="MM", rescale="MM", base.grp = 2)
tse_mm_N <- equate(aux_mm_N, D=1.7, method="TSE", base.grp = 2)
colnames(tse_mm_N) <- c("theta","SARA","NESSCA_MM")
```

Comparação das estimativas com o escore observado (Exemplo: NESSCA)

```
tse1_N <- merge(tse_mm_N[,c("SARA", "NESSCA_MM")], tse_ms_N[,c("SARA", "NESSCA_MS")], by = "SARA")
tse2_N <- merge(tse_hb_N[,c("SARA", "NESSCA_HB")], tse_sl_N[,c("SARA", "NESSCA_SL")], by = "SARA")
tse_N <- merge(tse1_N[,c("SARA", "NESSCA_MM", "NESSCA_MS")], tse2_N[,c("SARA", "NESSCA_HB", "NESSCA_SL")], by = "SARA")
valida_tse_N <- merge(valida[,c("Fonte", "id_bdorig", "SARA", "NESSCA")], tse_N[,c("SARA", "NESSCA_MM", "NESSCA_MS", "NESSCA_HB", "NESSCA_SL" )], by = "SARA")
```

Apuração das diferenças (Exemplo: NESSCA)

```
valida_tse_N$dif_mm <- abs(valida_tse_N$NESSCA_MM-valida_tse_N$NESSCA)
valida_tse_N$dif_ms <- abs(valida_tse_N$NESSCA_MS-valida_tse_N$NESSCA)
valida_tse_N$dif_hb <- abs(valida_tse_N$NESSCA_HB-valida_tse_N$NESSCA)
valida_tse_N$dif_sl <- abs(valida_tse_N$NESSCA_SL-valida_tse_N$NESSCA)
```

Gráfico Bland-Altman (Exemplo: NESSCA – Método *Mean/Mean*)

```
mm_bland_plot.1_N <- blandr.draw(valida_tse_N$NESSCA_MM, valida_tse_N$NESSCA,plotTitle="Bland-Altman")
mm_bland_plot_N <- mm_bland_plot.1_N +
        ggplot2::coord_cartesian( ylim=c(-20,20)) +
        labs(title="(a) Método Mean/Mean",
        y="Diferenças",
        x="Médias") +
        theme(plot.title = element_text(hjust = 0))
```

**ANEXO E**

**Curvas Característica do Item e Curvas de Informação do Item para cada item da NESSCA**

Este anexo apresenta as curvas de categoria de resposta e curvas de informação do item para cada item da NESSCA. Para as curvas de informação do item delimitou-se o eixo y até 7,0 para auxiliar na interpretação e comparação.

1 – Ataxia de Marcha

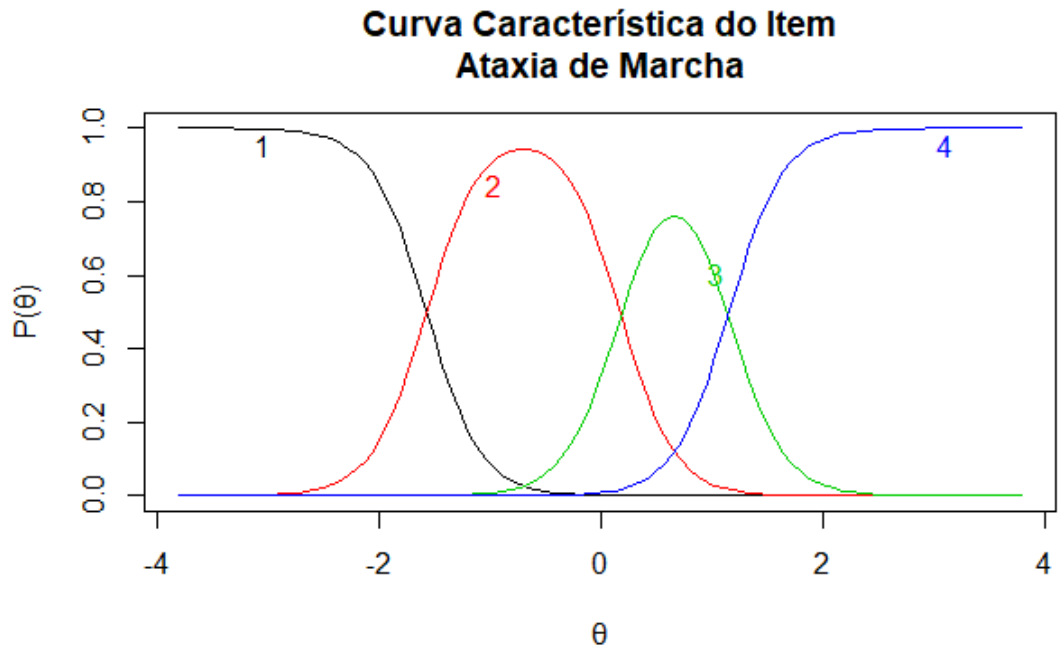

**Figura 2.** Curva Característica do Item para o item Ataxia de Marcha.

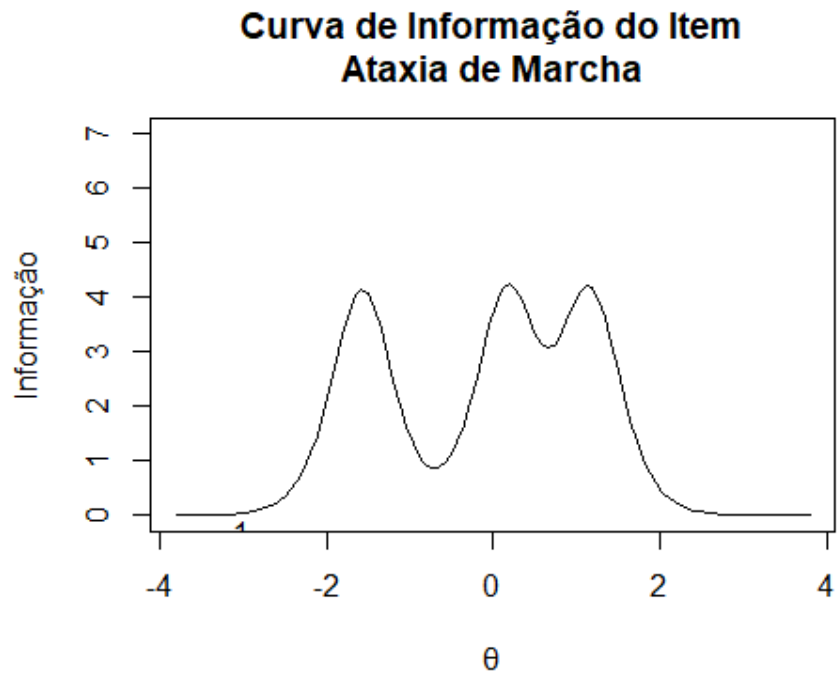

**Figura 3.** Curva de Informação do Item para o item Ataxia de Marcha.

2 – Ataxia nos Membros

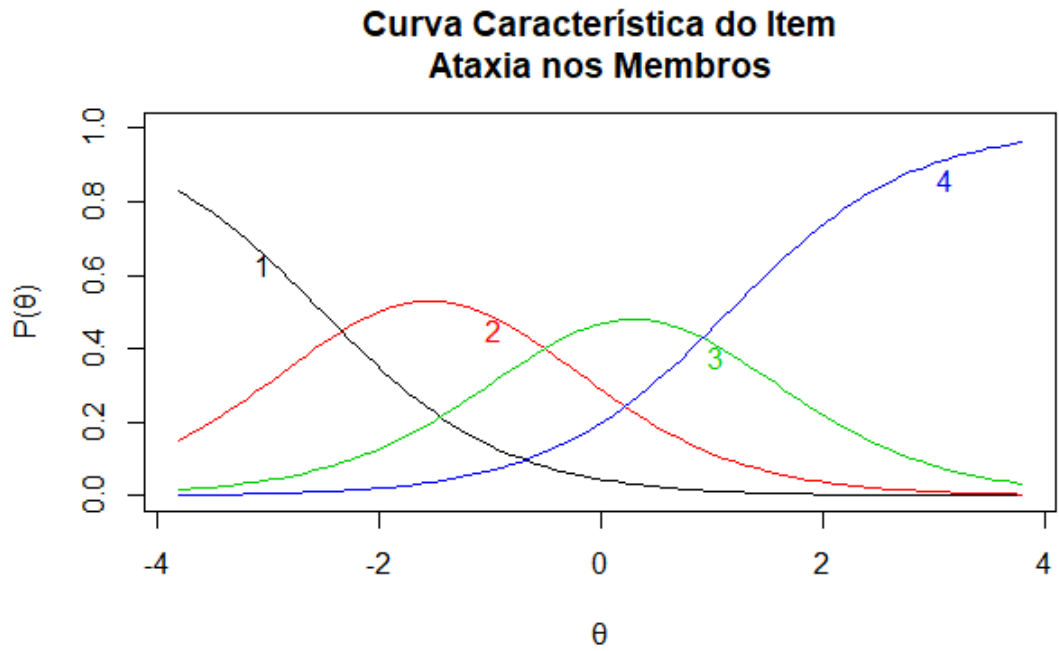

**Figura 4.** Curva Característica do Item para o item Ataxia nos Membros.

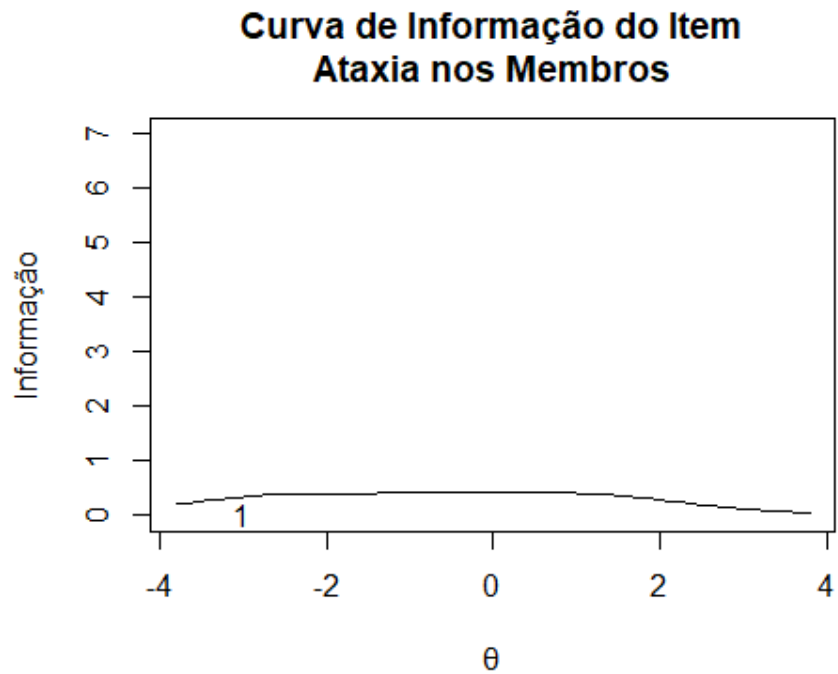

**Figura 5.** Curva de Informação do Item para o item Ataxia nos Membros.

3 – Nistagmo

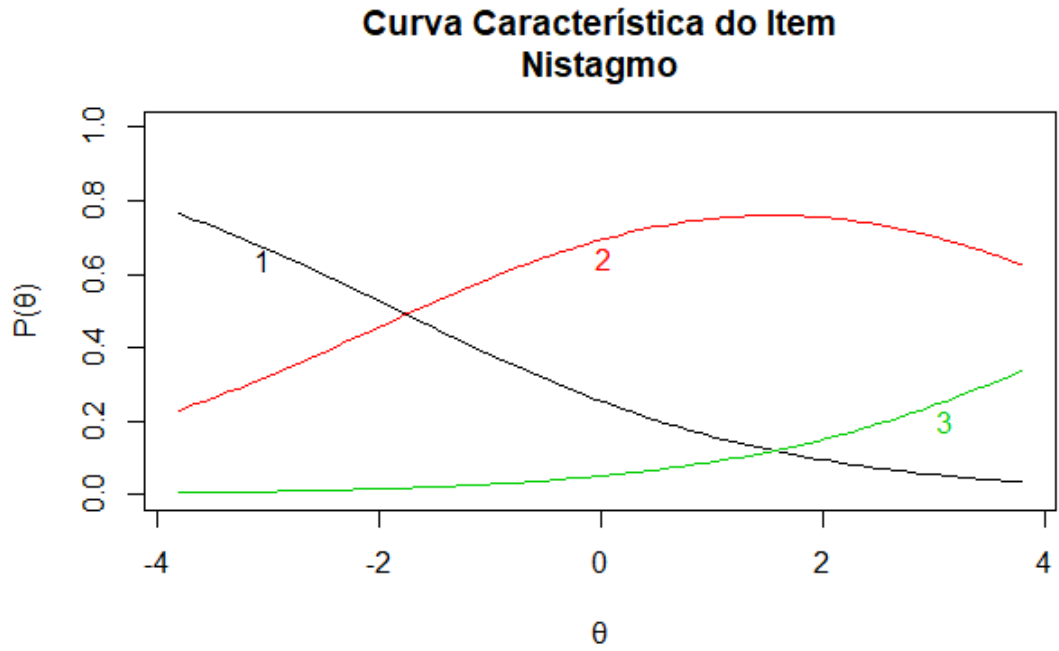

**Figura 6.** Curva Característica do Item para o item Nistagmo.

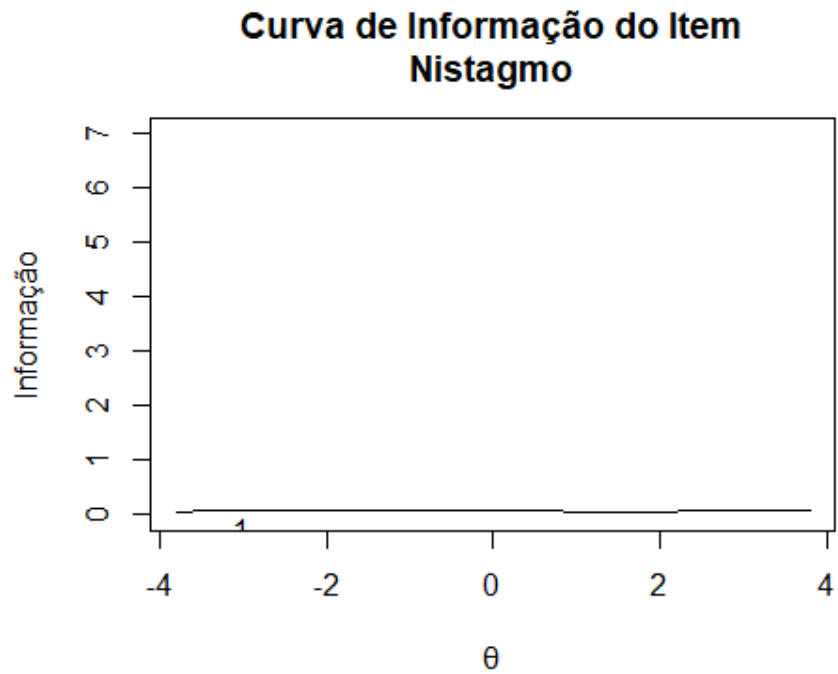

**Figura 7.** Curva de Informação do Item para o item Nistagmo.

4 – Disartria

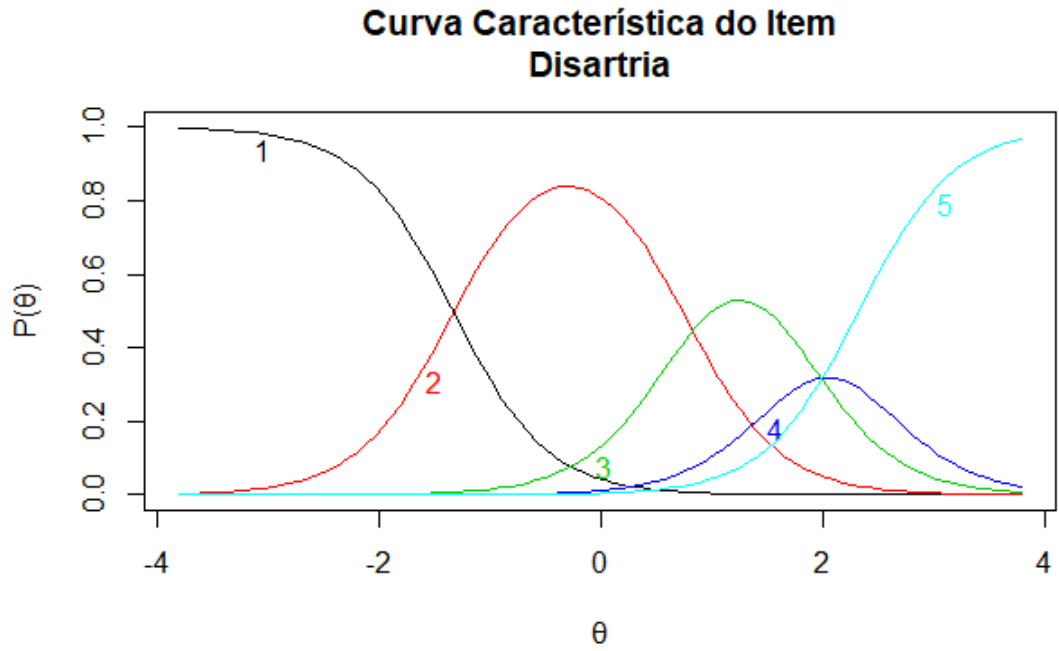

**Figura 12.** Curva Característica do Item para o item Disartria.

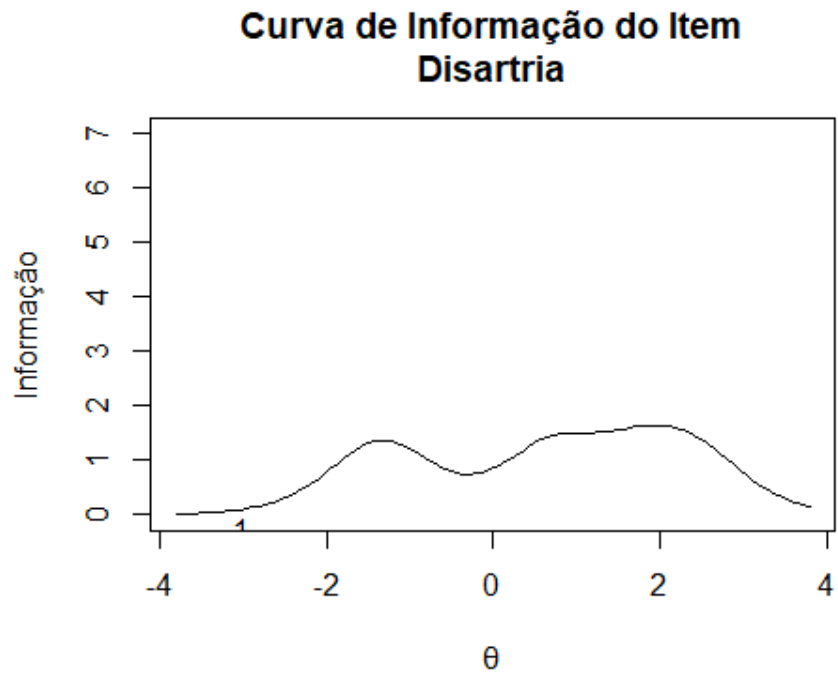

**Figura 13.** Curva de Informação do Item para o item Disartria.

5 – Disfagia

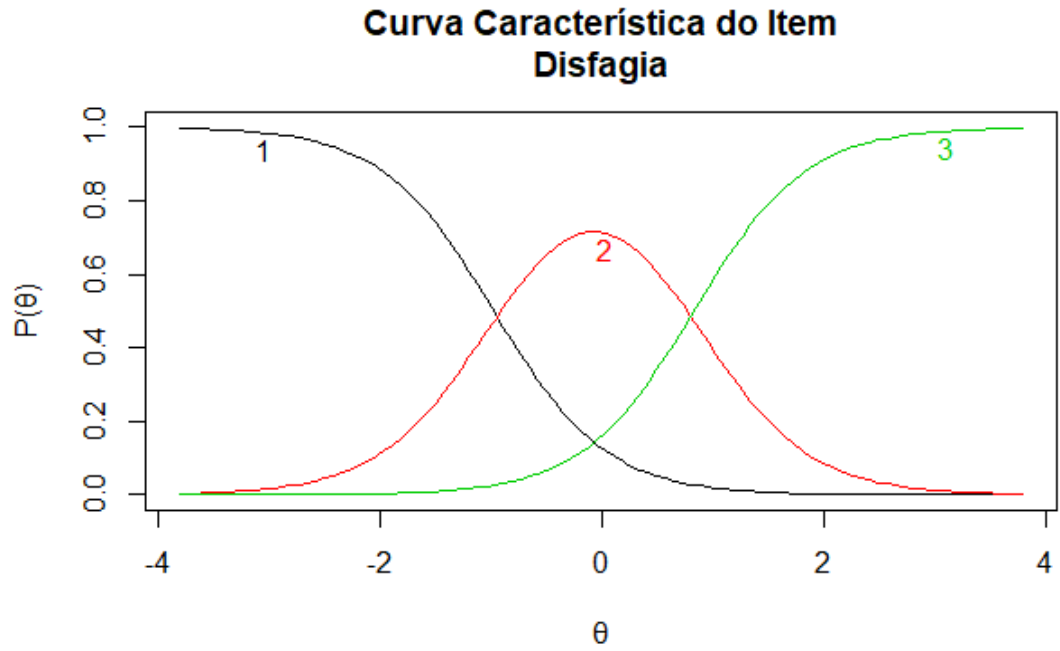

**Figura 14.** Curva Característica do Item para o item Disfagia.

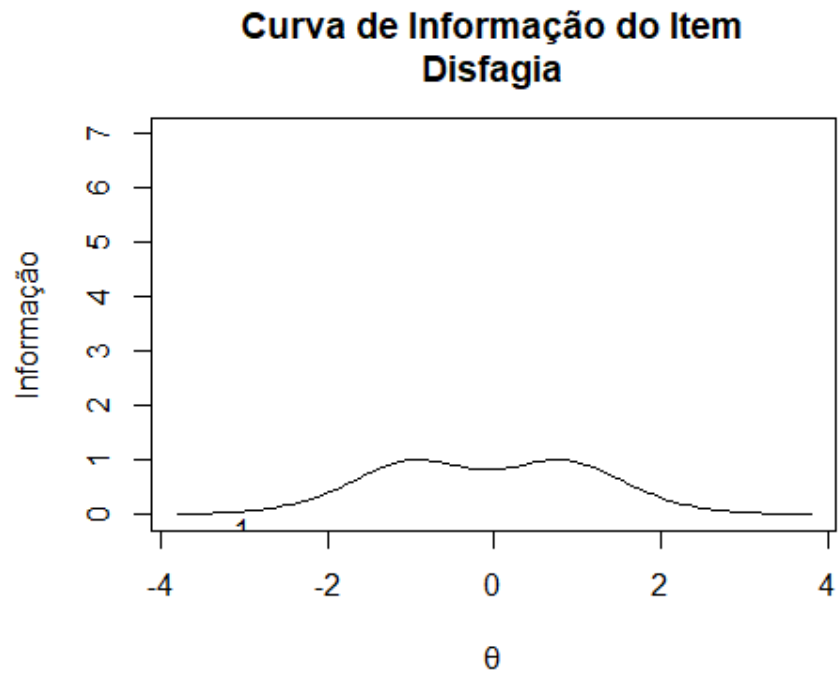

**Figura 15.** Curva de Informação do Item para o item Disfagia.

**ANEXO F**

**Curvas Característica do Item e Curvas de Informação do Item para cada item da SARA**

Este anexo apresenta as curvas de categoria de resposta e curvas de informação do item para cada item da SARA. Para as curvas de informação do item delimitou-se o eixo y até 7,0 para auxiliar na interpretação e comparação.

1 – Marcha

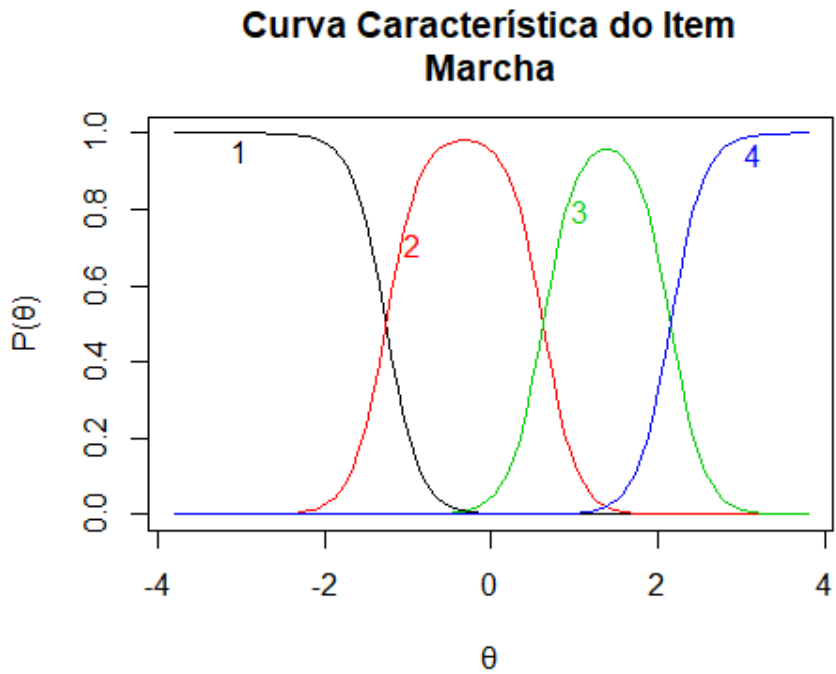

**Figura 38.** Curva Característica do Item para o item Marcha.

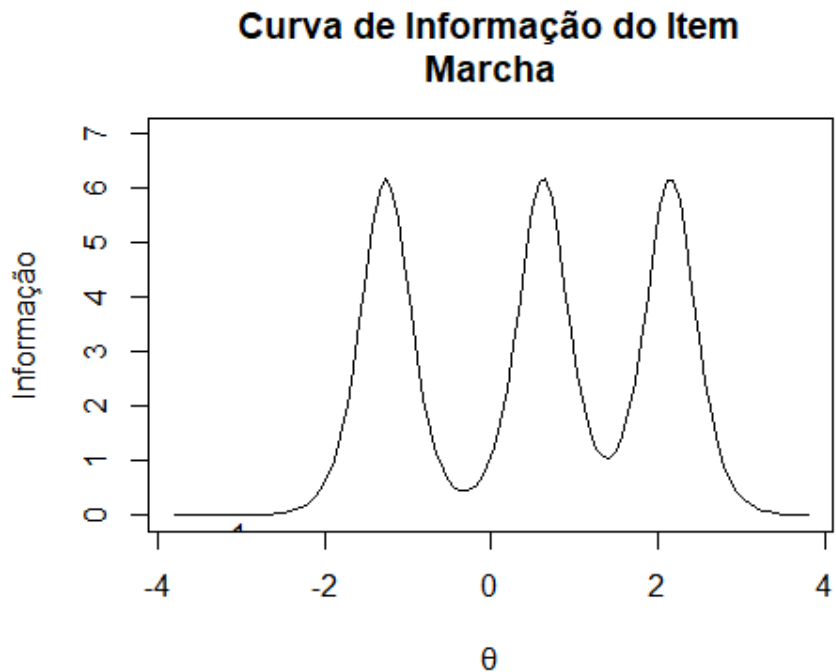

**Figura 39.** Curva de Informação do Item para o item Marcha.

2 – Equilíbrio de Pé

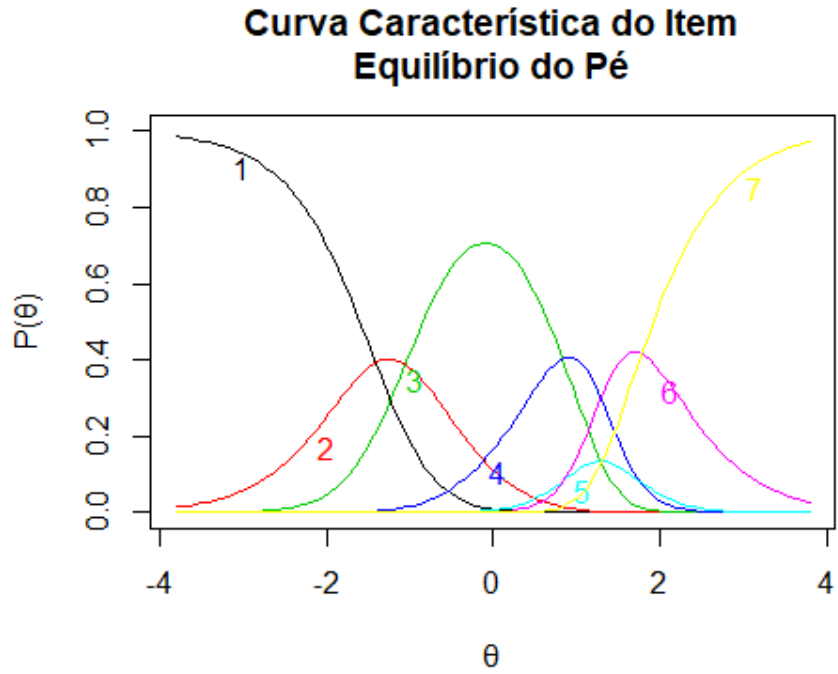

**Figura 40.** Curva Característica do Item para o item Equilíbrio de Pé.

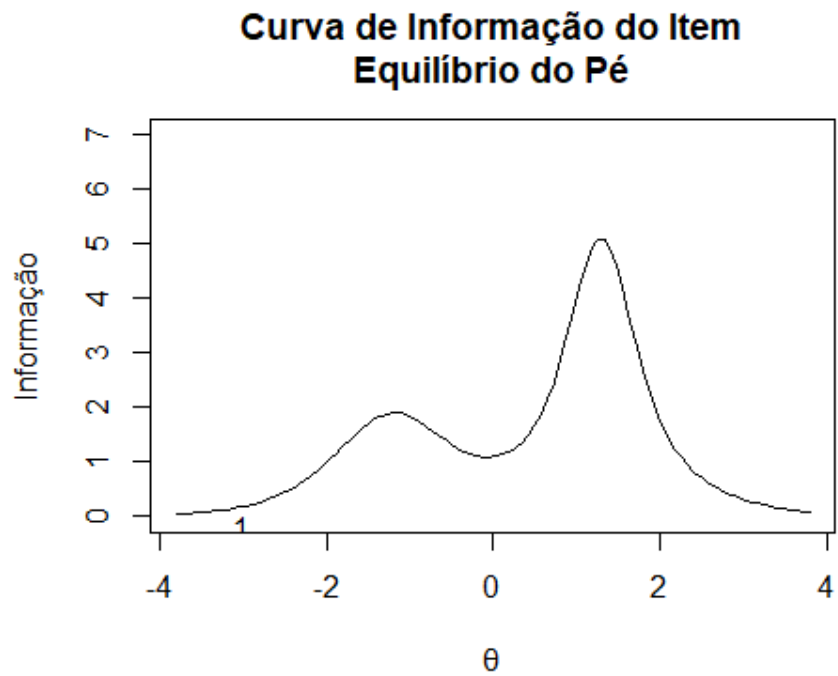

**Figura 41.** Curva de Informação do Item para o item Equilíbrio de Pé.

3 – Equilíbrio Sentado

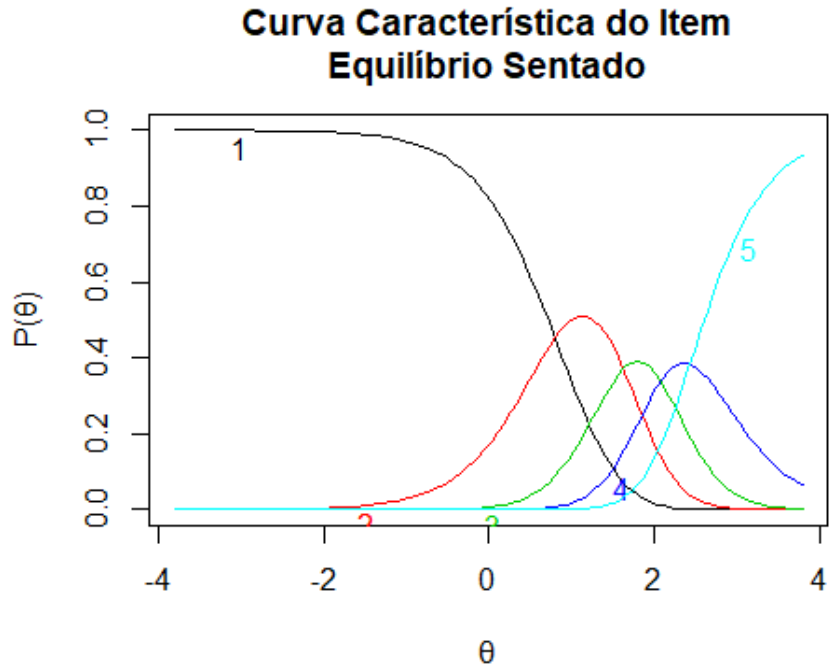

**Figura 42.** Curva Característica do Item para o item Equilíbrio Sentado.

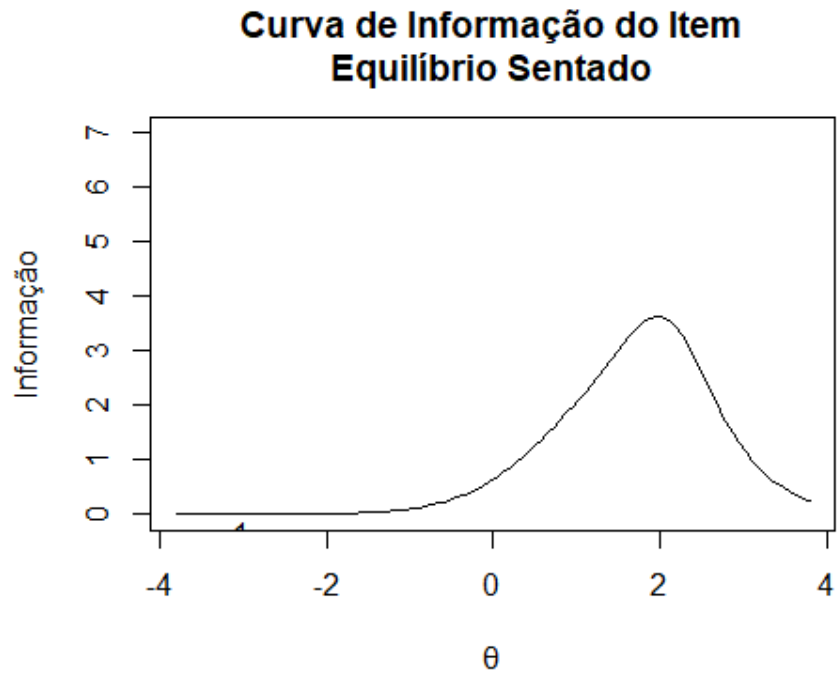

**Figura 43.** Curva de Informação do Item para o item Equilíbrio Sentado.

4 – Coordenação da Fala

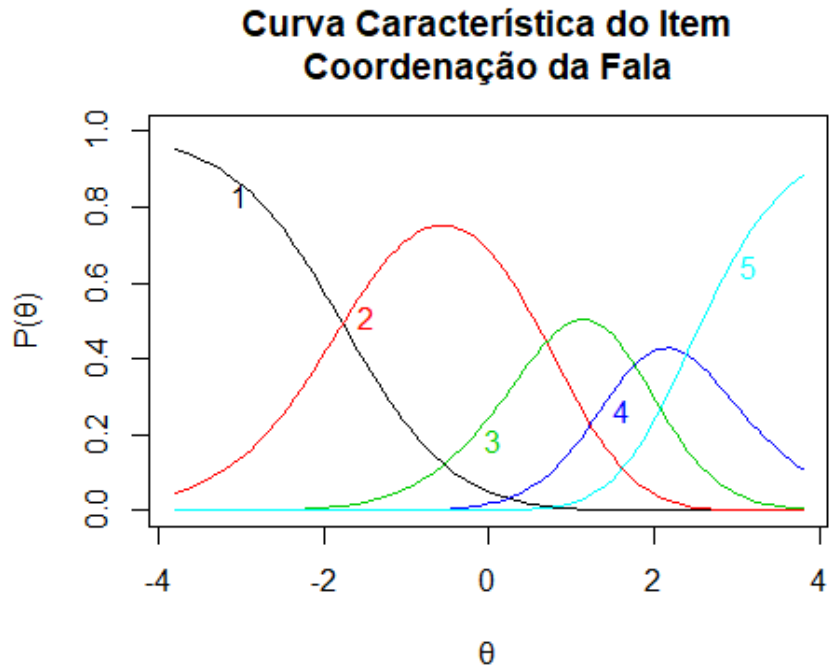

**Figura 44.** Curva Característica do Item para o item Coordenação da Fala.

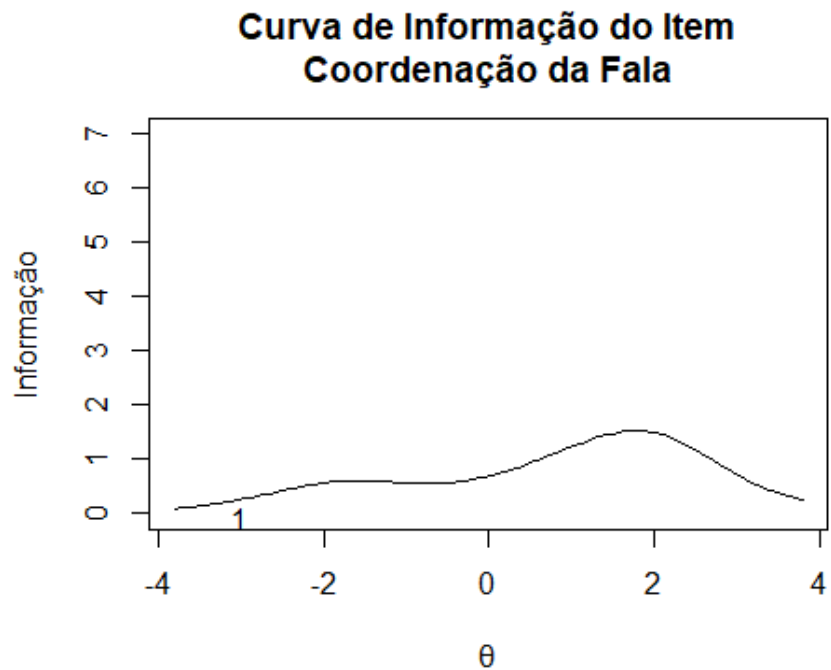

**Figura 45.** Curva de Informação do Item para o item Coordenação da Fala.

5 – Teste de Perseguição do Dedo

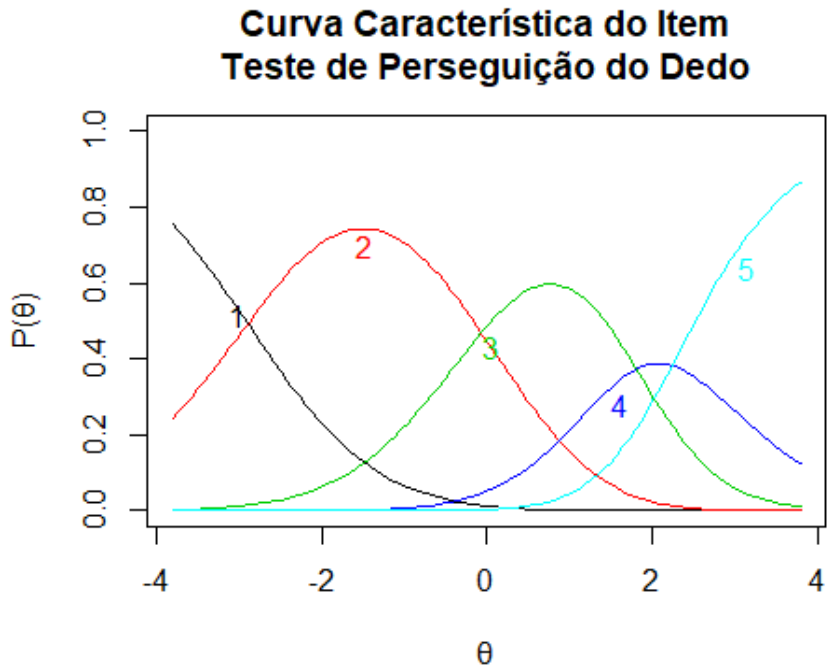

**Figura 46.** Curva Característica do Item para o item Teste de Perseguição do Dedo.

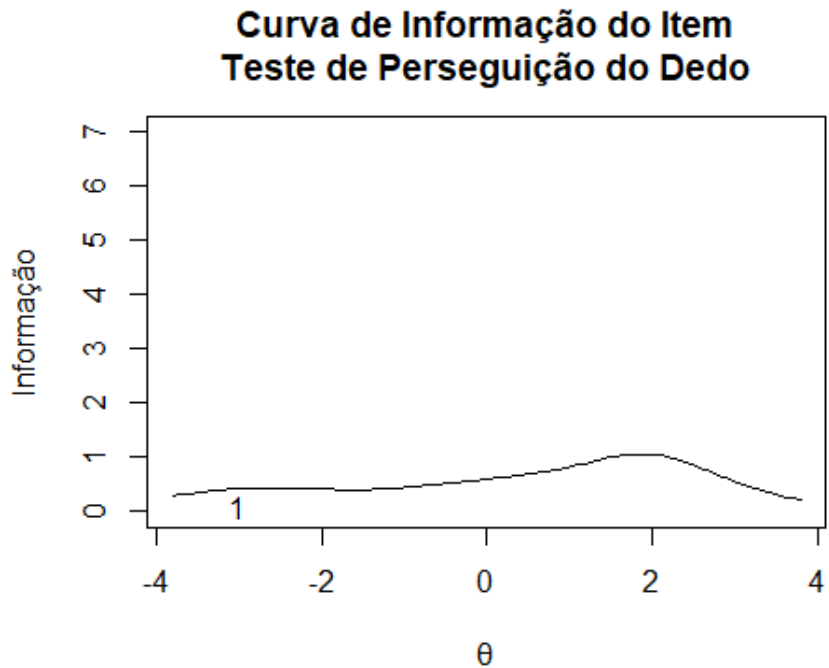

**Figura 47.** Curva de Informação do Item para o item Teste de Perseguição do Dedo.

6 – Teste Dedo-Nariz

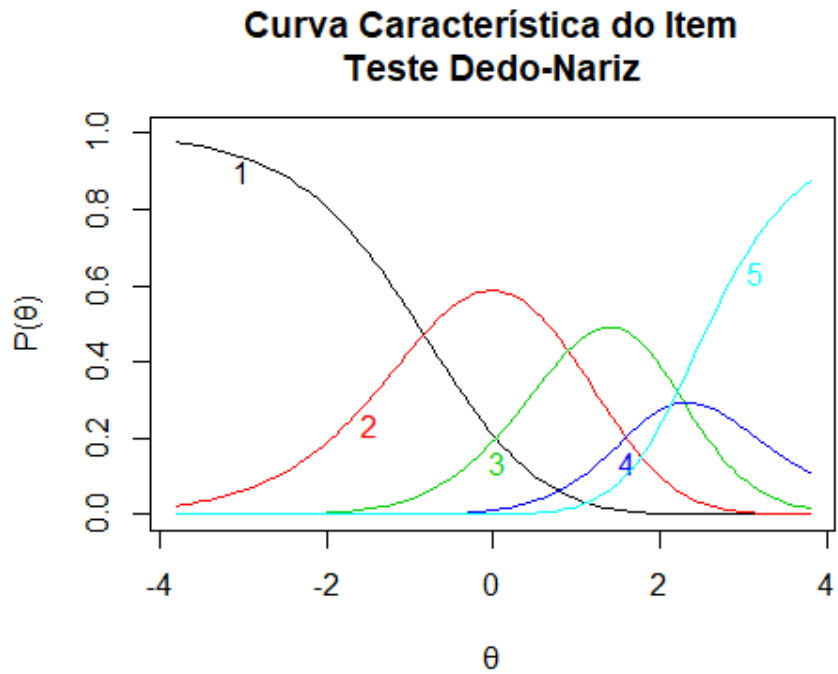

**Figura 48.** Curva Característica do Item para o item Teste Dedo-Nariz.

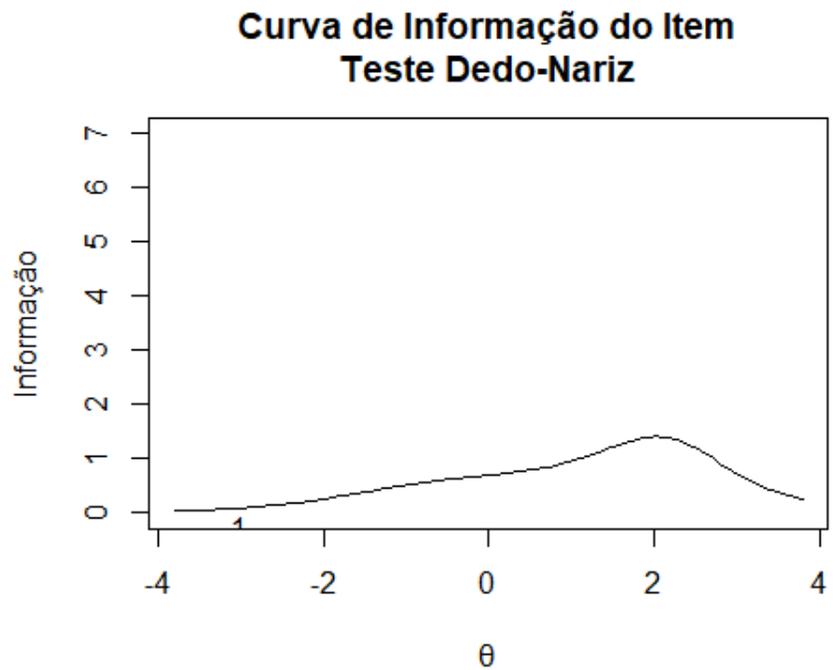

**Figura 49.** Curva de Informação do Item para o item Teste Dedo-Nariz.

7 – Diadococinesia

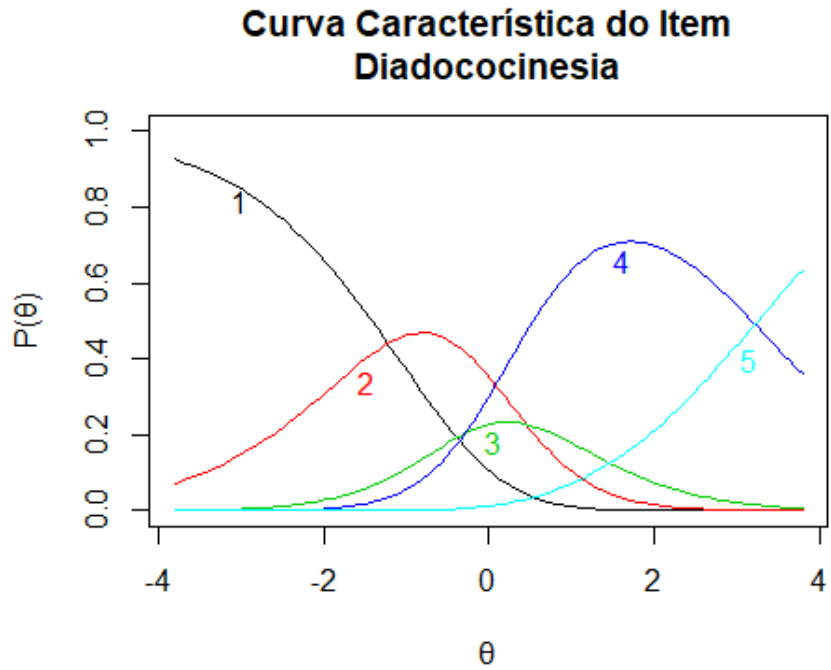

**Figura 50.** Curva Característica do Item para o item Diadococinesia.

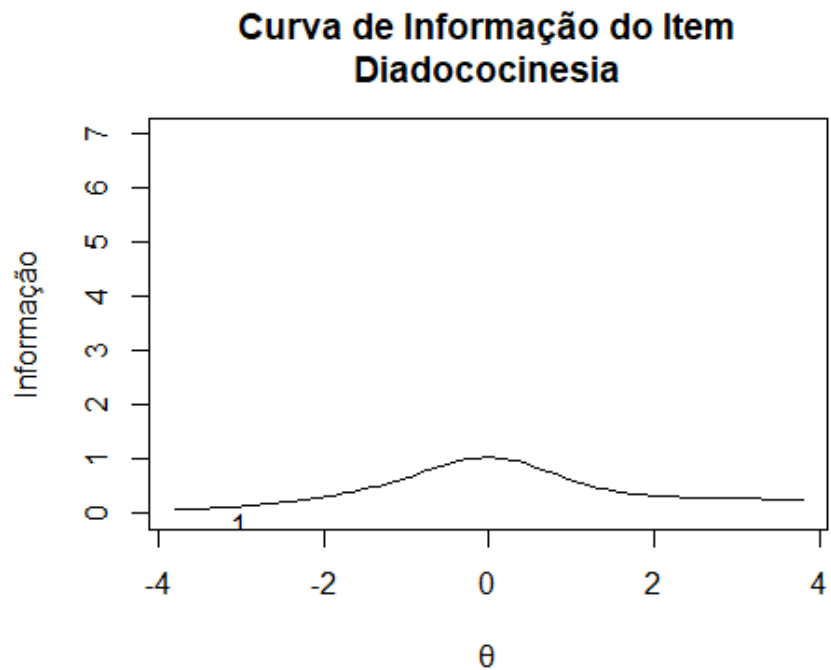

**Figura 51.** Curva de Informação do Item para o item Diadococinesia.

8 – Teste Calcanhar-Joelho-Canela

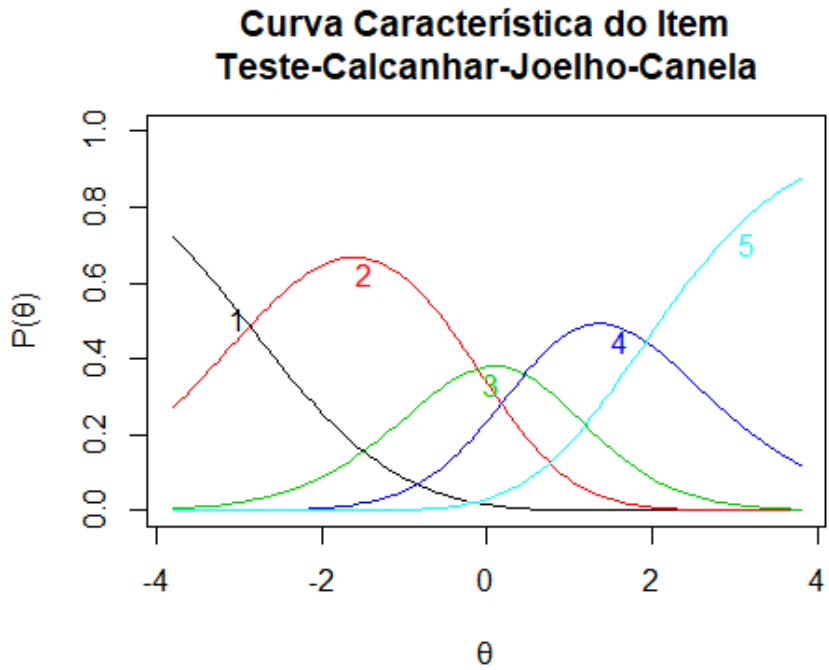

**Figura 52.** Curva Característica do Item para o item Teste Calcanhar-Joelho-Canela.

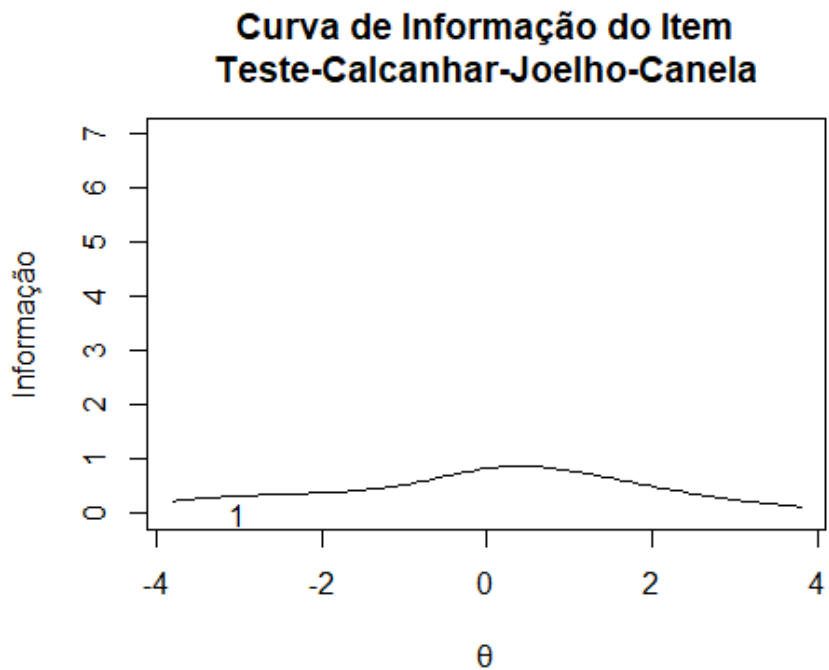

**Figura 53.** Curva de Informação do Item para o item Teste Calcanhar-Joelho-Canela.

**ANEXO G**

**Resultado da transformação linear**

**Tabela 1.** Resultado da transformação linear.

| Método de transformação | $\theta_N = A\theta_S + B$ | |
| --- | --- | --- |
| | **A** | **B** |
| *Mean/Mean* | 0,9803 | 0,2675 |
| *Mean/Sigma* | 1,1614 | 0,1951 |
| *Haebara* | 1,2354 | 0,4814 |
| *Stocking-Lord* | 1,1576 | 0,2301 |